\documentclass[useAMS,usenatbib,usegraphicx]{mn2e}
\usepackage{float,color}
\usepackage{epsf,rotating,url}
\usepackage{amssymb,amsfonts,amstext,amsgen,amsopn,amsxtra,indentfirst,lscape,times,rotating}
\usepackage{longtable,subfigure}
\usepackage{graphics}
\usepackage{keyval}
\usepackage{trig}
\usepackage[normalem]{ulem}
\usepackage{dcolumn}% Align table columns on decimal point
\usepackage{bm}% bold math
\def\beq{\begin{equation}}
\def\eeq{\end{equation}}
\def\bey{\begin{eqnarray}}
\def\eey{\end{eqnarray}}
\def\msun{M_\odot}
\def\lsun{L_\odot}
\def\kms{\, {\rm km \, s}^{-1} }

\def\grad{{\vec \nabla}}

\def\mnras{MNRAS}
\def\apj{ApJ}

\def\apjl{ApJ}

\def\araa{ARAA}

\def\aap{A \& A}
\def\aj{AJ}

\def\aap{Astron. Astrophys.}
\def\ser{S\'ersic }
\title{Mass models of disk galaxies from the DiskMass Survey in MOND}
\author[G. W. Angus, G. Gentile, R. A. Swaters, B. Famaey, A. Diaferio, S. S. McGaugh, K. J. van der Heyden]{
\Large G. W. Angus$^{1}$\thanks{E-mail: garry.angus@vub.ac.be}, G. Gentile$^{2,1}$,R. Swaters$^3$, B. Famaey$^4$, A. Diaferio$^{5,6}$, S. S. McGaugh$^7$, K. J. van der Heyden$^8$\\ 
$^{1}$Department of Physics and Astrophysics, Vrije Universiteit Brussel, Pleinlaan 2, 1050 Brussels, Belgium \\
$^{2}$Sterrenkundig Observatorium, Universiteit Gent, Krijgslaan 281, 9000, Gent, Belgium\\
$^{3}$National Optical Astronomy Observatory, 950 North Cherry Avenue, Tucson, AZ 85719, USA \\
$^{4}$Observatoire Astronomique de Strasbourg, Universit\'e de Strasbourg, CNRS UMR 7550, 11 Rue de l`Universit\'e, 67000 Strasbourg, France\\
$^{5}$Dipartimento di Fisica, Universit\`a di Torino, Via P. Giuria 1, I-10125, Torino, Italy \\
$^{6}$Istituto Nazionale di Fisica Nucleare, Via P. Giuria 1, I-10125, Torino, Italy\\
$^{7}$Department of Astronomy, Case Western Reserve University, 10090 Euclid Ave, Cleveland, OH 44106, USA\\
$^{8}$Astrophysics, Cosmology \& Gravity Centre, Dept. of Astronomy, University of Cape Town, Private Bag X3, Rondebosch, 7701, South Africa \\
}
\begin{document}
\date{\today}
\maketitle
\begin{abstract}
This article explores the agreement between the predictions of Modified Newtonian Dynamics (MOND) and the rotation curves and stellar velocity dispersion profiles measured by the DiskMass Survey. A bulge-disk decomposition was made for each of the thirty published galaxies, and a MOND Poisson solver was used to simultaneously compute, from the baryonic mass distributions, model rotation curves and vertical velocity dispersion profiles, which were compared to the measured values. The two main free parameters, the stellar disk's mass-to-light ratio ($M/L$) and its exponential scale-height ($h_z$), were estimated by Markov Chain Monte Carlo modelling. The average best-fit K-band stellar mass-to-light ratio was $M/L \simeq 0.55 \pm 0.15$. However, to match the DiskMass Survey data, the vertical scale-heights would have to be in the range $h_z=200$ to $400$~pc which is a factor of two lower than those derived from observations of edge-on galaxies with a similar scale-length. The reason is that modified gravity versions of MOND characteristically require a larger $M/L$ to fit the rotation curve in the absence of dark matter and therefore predict a stronger vertical gravitational field than Newtonian models. It was found that changing the MOND acceleration parameter, the shape of the velocity dispersion ellipsoid, the adopted vertical distribution of stars, as well as the galaxy inclination, within any realistic range, all had little impact on these results.
\end{abstract}
\section{Introduction}
\protect\label{sec:intr}
Understanding the dynamics of disk galaxies is essential to the vetting process of theories of galaxy formation and cosmology (\citealt{flores94,deblok98,deblok01,vandenbosch01,swaters03,gentile04,gilmore07b,deblok10}). Disk galaxies moderately inclined to the line of sight (50-80$^{\circ}$) can provide an HI rotation curve from which the dark matter (DM) content and distribution can be deduced in the context of Newtonian dynamics (\citealt{bosma78,rubin78,bosma81a,bosma81b,sofue01}). However, there exist degeneracies between the DM halo density profile, the stellar mass-to-light ratios of the stellar components and the scale-height of the stellar disk (\citealt{vanalbada85,kuijken89,kuijken91,angus12}).

The vertical stellar distribution is assumed to be a declining exponential on both sides of the mid-plane. Thus, the scale-height of the stellar disk is simply the exponential rate at which the stellar luminosity density drops, with increasing height above or below the mid-plane of the disk. It is assumed to be constant with radius, which is supported by observations (e.g. \citealt{vanderkruit81a,vanderkruit82,bizyaev02}).

The stellar mass-to-light ratio ($M/L$) is the ratio between the mass of a stellar population and the luminosity, as observed through a particular bandpass. It cannot be observed directly, so it is ordinarily inferred by numerical stellar population synthesis models. This modelling crucially depends on the star-formation and chemical enrichment history considered and on the initial mass function (IMF; \citealt{belldejong,chabrier03,kroupa01}). The IMF is the spectrum of stellar types formed given a molecular cloud of certain initial mass, metallicity and other relevant properties. It therefore has the scope to vary from galaxy to galaxy. In addition to the IMF, stellar population synthesis models have other sources of uncertainty (\citealt{conroy09,conroy10a,conroy10b}).

In standard fitting of rotation curves of highly inclined disk galaxies, it is common to invoke the ``maximum disk hypothesis'', (\citealt{vanalbada86,sackett97,courteau99}) i.e. that the stellar disk contributes maximally to the rotation curve.
This hypothesis is supported by observations that deduce the microlensing optical depth in the Milky Way (e.g., \citealt{bissantz02}), the baryonic Tully-Fisher relation (\citealt{mcgaugh15}) and measurements that place the co-rotation radius of barred galaxies just beyond the end of the bar (e.g., \citealt{sellwood14}). It yields values for the $M/L$ that are typically in accordance with the predictions of stellar population synthesis models.

This, however, does not demonstrate that the hypothesis is correct (see e.g. \citealt{herrmann09,dutton11}) and it would be ideal to have a robust, independent measurement that breaks the disk mass degeneracy. This is theoretically possible because the $M/L$ can be determined dynamically through the vertical velocity dispersions of stars over the full projected area of the galactic disk, if the scale-height is known (\citealt{bahcall84}). For close to edge-on disk galaxies, however, one cannot measure the vertical velocity dispersion and thus the technique is limited to only those galaxies with moderate inclinations to the line of sight ($5-45^{\circ}$), where $25<i<35^{\circ}$ is seen as optimal (\citealt{bershady10b}; hereafter DMSii). 

In order to break the degeneracy between the DM halo and $M/L$, the DiskMass Survey (DMS; \citealt{bershady10a}; hereafter DMSi) made observations of the line of sight velocity dispersion profiles of 46 nearly face-on disk galaxies (DMSi):  30 of which have been published and a further 100 are part of the larger survey. They also measured their surface brightness profiles and rotation curves. There were two further, essential ingredients in the analysis that come from scaling relations. The first is the luminous Tully-Fisher (TF) relation between the absolute magnitude of a galaxy and a measure of its outer rotation speed. This helps to isolate the inclination of the galaxy, which for nearly face-on galaxies can otherwise be obtained using the tilted disk method of \cite{andersen13}.

The second is the correlation between the disk stellar scale-height, a measure of the thickness of the disk, and scale-length, a measure of the radial extent. This relationship is explored in detail by  DMSii (their \S2.2), and is derived from the studies of \cite{kregel02} (hereafter K02) and \cite{pohlen00,schwarzkopf00,xilouris97,xilouris99}. This allowed \cite{bershady11} and \cite{martinsson13b,martinsson13a} (hereafter DMSvi and DMSvii) to infer the $M/L$ of the disk. The data imply that the stellar disks are  ``sub-maximal'' (K-band $M/L \simeq 0.3$ or lower, see \citealt{swaters14}) which means they do {\it not} contribute maximally to the rotation curve in the central regions, contrary to the value found from population synthesis models that assume a Kroupa IMF ($M/L \simeq 0.6$; \citealt{mcgaugh14}). This leaves more room for DM in the central regions.

Modified Newtonian Dynamics (MOND \citealt{milgrom83a}, see \citealt{famaey12} for a recent review) is a theory which proposes a modification of dynamics whose impact is most apparent in regions of low acceleration. Most current working versions of MOND consist of an actual modification of gravity, i.e., at the classical level, a modification of the Newtonian Poisson equation (but see also \citealt{milgrom11}). This modification occurs due to the hypothesised existence of a new constant of physics with dimensions of acceleration, $a_0\sim10^{-10}\rm ms^{-2}$. For accelerations much stronger than this threshold, $a_0$, there is no discerned deviation from Newtonian gravity. However, far below the threshold the true acceleration perceived by a test mass is found from $a^2=g_Na_0$ - where $g_N$ is the expected Newtonian gravitational field.

\cite{nipoti07b} and \cite{bienayme09} made studies of the vertical dynamics of the Galaxy in MOND, showing that it could be possible to distinguish between MOND and the DM paradigm with data from the Milky Way. The extra constraint on the dynamics from vertical velocity dispersions in nearly face-on disks of external galaxies provides a new test of this hypothesis. This article addresses whether MOND can simultaneously account for the measured vertical velocity dispersions and rotation curves, while keeping in line with galaxy scaling relations.

In section 2 the framework is presented for the joint modelling of galaxy rotation curves and stellar vertical velocity dispersions in the MOND context. In section 3 the methods are discussed and this includes a discussion of the bulge-disk decomposition, the accuracy of the Poisson solver, and the observational error budget for the main data. In section 4 the primary results are presented, this includes a discussion of the fits to the vertical dynamics and rotation curves, the confidence ranges of the fitted parameters, and how well the fitted parameters mesh with other observations. In section 5 possible scenarios that could alter the results are discussed. In section 6 conclusions are drawn and their implications are explored.

\section{Dynamical analysis of the DiskMass survey}
\protect\label{sec:dyn}
\subsection{Rotation curve fitting}
\protect\label{sec:rcf}
The following reviews how to fit the measured rotation curve of a disk galaxy in an idealised case. This is done in order to expose the free parameters and it is generalised for both DM and MOND. 

The total model rotation speed is required from the total model potential in order to compare with the measured rotation curve. This can be found from

\beq
\protect\label{eqn:vrot}
{V_{tot}(R)^2 \over R}= {d\Phi_{tot} \over dR}(R).
\eeq
Here, $R$ is the cylindrical radius in the disk mid-plane and $\Phi_{tot}$ is the total gravitational potential. Next, the total potential is required from the total mass distribution. This is computed via the Poisson equation, which in Newtonian dynamics is

\beq
\protect\label{eqn:newtpois}
\grad^2\Phi_{n,tot}(R,z)=4\pi G\rho_{tot}(R,z),
\eeq
where $G$ is Newton's gravitational constant, $\rho_{tot}(R,z)$ is the total mass density from all sources (see the following and \S\ref{sec:bardens}) and $\Phi_{n,tot}(R,z)$ is the total {\it Newtonian} potential. In Newtonian dynamics, $\Phi_{n,tot}(R,z)$ is fully equivalent to $\Phi_{tot}(R,z)$, but $\rho_{tot}$ consists of $\rho_{bar}$ and $\rho_{DM}$. In MOND, $\rho_{tot}$ is fully equivalent to $\rho_{bar}$, but a second step is made to find $\Phi_{tot}$ from $\Phi_{n,tot}$, which is (\citealt{milgrom10})

\beq
\protect\label{eqn:mondpois}
\grad^2\Phi_{tot}(R,z)=\grad \cdot \left[ \nu(|\grad\Phi_{n,tot}| / a_0) \grad\Phi_{n,tot} \right],
\eeq
where $a_0\sim3.6(\kms)^2\rm pc^{-1}$ is the acceleration threshold of MOND and $\nu$ is an interpolating function, chosen here to depend on its argument as
\beq
\protect\label{eqn:nualp}
\nu_{\gamma}(y)=\left[{1+(1+4y^{-\gamma})^{1/2}\over 2}\right]^{1/\gamma},
\eeq
where $\gamma=1$ is the simple $\nu$-function and $\gamma=2$ is the standard $\nu$-function (\citealt{famaey12} Eqs. 51 and 53), or likewise by
\beq
\protect\label{eqn:nudel}
\nu_{\delta}(y)=\left[1-e^{-y^{\delta/2}}\right]^{-1/\delta}+(1-\delta^{-1})e^{-y^{\delta/2}}.
\eeq

\subsubsection{Baryonic density}
\protect\label{sec:bardens}
The last question here is the distribution of $\rho_{tot}$. The derivation of the contribution from atomic and molecular gas is described in DMSvii. Both are assumed to have non-smooth, axisymmetric, radial surface densities. Both gas components are included in the modelling, but they are considered fixed in mass and distribution. They are given nominal scale-heights of 200~pc, to which reasonable variations are inconsequential.

There are typically two stellar components, a bulge and a disk. Below, the bulge and disk surface brightnesses are given to expose the free parameters in their fitting. The bulge is assumed to be spherical and to follow a \ser profile
\beq
\protect\label{eqn:sersic}
I_b(R)=I_e\exp\{-k[(R/R_e)^{1/n}-1]\},
\eeq
where the effective surface brightness ($I_e$), the effective projected radius ($R_e$) and the \ser index ($n$) can be fitted to the observed surface brightness distribution.\footnote{The constant $k$ is fixed at 7.67.} Simultaneously, the disk luminosity density can be expressed with a simple form

\beq
\protect\label{eqn:jdisk}
j_{d}(R,z)={L_{d} \over 4\pi h_R^2h_z}\exp\left(-{R \over h_R}\right)\exp\left(-{|z| \over h_z}\right),
\eeq
where $L_{d}$ is the total luminosity of the disk, $h_R$ is the scale-length and $h_z$ is the scale-height. The vertical distribution is characterised by the exponential function, but the commonly used $sech^2(z/z_0)$ function (\citealt{vanderkruit81a,bottema93}) would be equally appropriate. Here $z_0$ is the $sech^2$ scale-height which corresponds to $2h_z$ at large $z$. Using the $sech^2$ vertical distribution does not change the conclusions (see \S\ref{sec:sech}).

In general, the surface density can be found by integrating along the line of sight. For a face-on galaxy, the luminosity density of Eq~\ref{eqn:jdisk} can be projected to give the surface brightness

\beq
\protect\label{eqn:sigdisk}
I_{d}(R)={L_{d} \over 2\pi h_R^2}\exp\left(-{R \over h_R}\right).
\eeq
Here the density profiles are always assumed to be smooth. To find the mass density of the bulge and disk, the luminosity densities of the bulge and the disk must be multiplied by their respective $M/L$ so that

\beq
\protect\label{eqn:massdens}
\rho_*=\Upsilon_{b}j_b+\Upsilon_{d}j_d,
\eeq
where $\Upsilon_{b}$ and $\Upsilon_{d}$ are the stellar $M/L$ values of the bulge and disk respectively.
The total density of baryons is then $\rho_{bar}=\rho_*+\rho_g$, where $\rho_g$ is the atomic and molecular gas density. As stated previously, in MOND, $\rho_{bar}$ is equivalent to the total mass density, $\rho_{tot}$ (Eq~\ref{eqn:newtpois}), because there is no DM in MOND galaxies. For Newtonian gravity $\rho_{tot}=\rho_{bar}+\rho_{DM}$.

For the baryonic mass models, there are 8 parameters: $\Upsilon_{b},I_e,R_e,n,L_d,\Upsilon_{d},h_R,h_z$. Of those 9, the surface brightness parameters are either directly observed or unambiguously fitted to the surface brightness profile. This leaves only $\Upsilon_{b},\Upsilon_{d},h_z$. Since the DMS sample is chosen so that the total bulge luminosity to total disk luminosity is low, $\Upsilon_{b}$ is relatively insignificant and is never independently varied in the modelling performed here ($\Upsilon_{b}=\Upsilon_{d}$). Therefore, $\Upsilon_d$ and $h_z$ are the only two free parameters that are relevant to a theory like MOND. In principle, the inclination of the galaxy also has a small amount of freedom but it is strongly curtailed by the luminous TF relation. The aforementioned free parameters are fitted for through a simultaneous comparison of the model vertical velocity dispersions and rotation curves with the observed ones, as is described in \S\ref{sec:errors}.

\subsubsection{Inclination}
\protect\label{sec:inc}
The derivation of the rotation curve of a moderate or high-inclination disk galaxy from the measured 2D velocity field, permits the fitting of tilted rings (see \citealt{begeman89,vdhulst92}). These tilted rings allow us to model the variation in the inclination and position angle of numerous concentric annuli at different galactocentric radii. The inclinations of the various rings, as a function of radius, can vary by 10$^{\circ}$ (e.g. \citealt{deblok08}) depending on the quality of the data, the regularity of the velocity field and characteristic inclination of the disk.

The DMS galaxies have low inclinations (they are close to face-on), thus the inclination has a lot of leverage on the inferred rotation speed because of the shape of the sine function. It is possible to derive accurate kinematic inclinations for nearly face-on disks using the tilted disk (as opposed to tilted ring) technique of \cite{andersen13}.  It is also possible to infer the inclination using the luminous TF relation (\citealt{verheijen01}). This relates the absolute K-band magnitude of the galaxy, $M_K$, to a measure of the outer rotation speed, $V_f$ such that 
\beq
\protect\label{eqn:tfr}
V_f=0.5 \times 10^{(5.12-M_K)/11.3}~\kms.
\eeq
Anderson et al. (2015, in prep) have determined that the kinematic and TF inclinations for the DiskMass galaxies generally agree well, although there are some outliers.

Relating the measured, inclined outer rotation velocity $V_{obs}sin(i)$ with the expected outer velocity from the TF relation (Eq~\ref{eqn:tfr}) allows the expected inclination to be deduced. This inclination is only that expected for the outer parts of the rotation curve, and thus the inner parts can vary somewhat due to a warp. 

In addition to the Luminous TF relation, there is a baryonic TF relation (\citealt{mcgaugh00,mcgaugh05a}) which relates the total baryonic mass of a galaxy to its outer, flat rotation speed, where $V_f^{\epsilon} \propto GM_{bar}$.

This relation is fundamental to MOND and the exponent $\epsilon=4$ and the constant of proportionality $a_0$ are predictions which agrees well with the observed relation (\citealt{mcgaugh05a}). Thus, the MOND baryonic TF relation can be written in a similar form to Eq~\ref{eqn:tfr} as

\beq
\protect\label{eqn:btfr}
{V_f^4 \over Ga_0}=M_{g}+\left(\Upsilon_{b}f_b+\Upsilon_{d}f_d\right)\times10^{(M_{K,\odot}-M_K)/2.5}.
\eeq
Here, $f_b={L_b \over L_b+L_d}$ and $f_d$ are the fractions of the total luminosity contributed by the bulge and disk respectively. The absolute magnitude of the Sun in the K-band is $M_{K,\odot}=3.28$ (\citealt{blanton03}) and $M_g$ is the gas mass.

It was found that the inclinations from Eq~\ref{eqn:btfr} are typically between 5-15\% larger than those found with Eq~\ref{eqn:tfr}, depending on the $M/L$ used (here the $M/L$ was taken to be between 0.6 and 1): a smaller $M/L$ implies a smaller corresponding rotation velocity, hence a larger inclination.

For a large enough sample there should, in principle, be no systematic deviation from either TF relation. When modelling rotation curves in general, it is not always clear when the rotation curve has reached the terminal velocity, so some margin of error must be granted. Since Eq~\ref{eqn:tfr} has no dependence on $M/L$, and to make it easier to compare with the DMS results, the luminous TF relation inclinations (Eq~\ref{eqn:tfr}) are used in this article.

\subsection{Stellar vertical velocity dispersions}
\protect\label{sec:svvd}
\subsubsection{Choice of stellar velocity dispersion ellipsoid parameters}
\protect\label{sec:csve}

In addition to the measured rotation curve, the DMS also measured the line of sight velocity dispersion profile of the stars over the full projected area of the disk. This is then azimuthally averaged to give a 1D line of sight velocity dispersion, which can be converted to a vertical velocity dispersion ($\sigma_z$) through the equation (\citealt{westfall11})

\beq
\protect\label{eqn:vellip}
\sigma_z^2={\sigma_{los}^2 \over \cos^2i}\left[1+{\tan^2i \over 2\alpha^2}(1+\beta^2)\right]^{-1}.
\eeq
Here, the inclination of the galaxy to the line of sight is again, $i$, and $\alpha={\sigma_z \over \sigma_R}$ \& $\beta={\sigma_{\theta} \over \sigma_R}$ provide information about the stellar velocity dispersion ellipsoid. Generally, $\alpha$ and $\beta$ are expected to take on certain values from measurements in the Solar neighbourhood (\citealt{binneymerri98,gerssen12}), and $\beta$ is presumed to take on specific values from the epicycle approximation. However, beyond the Milky Way their variation is not empirically well known (see discussion in sect 2.1 of DMSii, and also \citealt{westfall11,westfall15,gentile15}). By choosing galaxies that are nearly face-on, the DMS reduces their importance (cf. Fig~\ref{fig:vvdvar}). The statistical variation of these parameters is discussed in DMSii \S2.1, and the DMS analysis establishes $\alpha=0.6\pm0.15$ and $\beta=0.7\pm0.04$. The mean values used by the DMS are chosen for the default values in this article. 
\subsubsection{Model Stellar vertical velocity dispersions}
\protect\label{sec:modsvvd}

In order to compare with the observations, the model vertical velocity dispersions of the galaxies must be computed. The vertical velocity dispersion at a height $z$ above the mid-plane, at a radial distance $R$ from the centre of the disk galaxy is found from (see \citealt{nipoti07b})

\beq
\protect\label{eqn:rho*}
\rho_*(R,z)\sigma_z(R,z)^2=\int_z^{\infty}\rho_*(R,z'){d\Phi_{tot}(R,z') \over dz'}dz',
\eeq
and the equivalent of the observed vertical velocity dispersion at any radius, R, weighted by the local stellar surface density is given by

\beq
\protect\label{eqn:sig*}
\Sigma_*(R)\sigma_z(R)^2=\int_{-\infty}^{\infty}\rho_*(R,z)\sigma_z(R,z)^2  dz.
\eeq
These equations effectively reduce to

\beq
\protect\label{eqn:sigred}
\sigma_z(R)^2={1 \over h_z}\int_{0}^{\infty}\left[ \int_z^{\infty}\exp(-z'/h_z){d\Phi_{tot}(R,z') \over dz'}dz' \right]dz.
\eeq

In the Newtonian gravity framework, Eq~\ref{eqn:sigred} depends mainly on 2 fitted parameters: $\Upsilon_{d}$ and $h_z$. As stated previously, inclination could also be varied but only in a tight range around the TF relation values. Thus, combining simultaneous fits to the observed rotation curves and vertical velocity dispersions is a strong test of the MOND paradigm.

It is worth noting that the infinite potential well of isolated galaxies in MOND is irrelevant here since the vertical gravitational field in Eq~\ref{eqn:rho*} is  convolved with the exponentially declining stellar density, and thus the MOND gravity is only relevant where there are stars.

\subsubsection{The DiskMass Survey Method}
\protect\label{sec:dmsmethod}
In the analysis of DMSvi, the measured rotation curve of each galaxy is used to fit the DM halo parameters and then the vertical velocity dispersion is used to directly give the mass surface density using $\Sigma_{dyn}(R)={\sigma_z(R)^2 \over \pi G k h_z}$, where $k$ is assumed to be 1.5 to describe an exponential vertical stellar distribution. From this surface density, the gas disk was subtracted. This left the stellar disk surface density, $\Sigma_*(R)=\Sigma_{dyn}(R)-\Sigma_{gas}(R)$. This also includes an unknown component of DM. The stellar $M/L$ as a function of radius was given by $\Upsilon_d(R)={\Sigma_*(R) \over I_d(R)}$. An average of this $M/L$ out to a given radius then defines the quoted disk $M/L$. This approach, although straight-forward, is only accurate when the derivative of the rotation curve is small. It is also not transferable to MOND because of the non-linearity of the theory and that the $M/L$ in MOND affects both the rotation curve and vertical velocity dispersion. It is more secure to make the reverse calculation and go from observed surface brightness, sample a $M/L$ to give surface density, then use Eq~\ref{eqn:sigred} to give a model vertical velocity dispersion which can be compared with the observed vertical velocity dispersions. However, this requires calculation of the full three dimensional potential from a model galaxy - which is performed here.

\begin{figure*}
\def\subfigtopskip{0pt} 
\def\subfigbottomskip{4pt}
\def\subfigcapskip{1pt}
\centering
\begin{tabular}{cc}
\subfigure{
\includegraphics[angle=0,width=8.0cm]{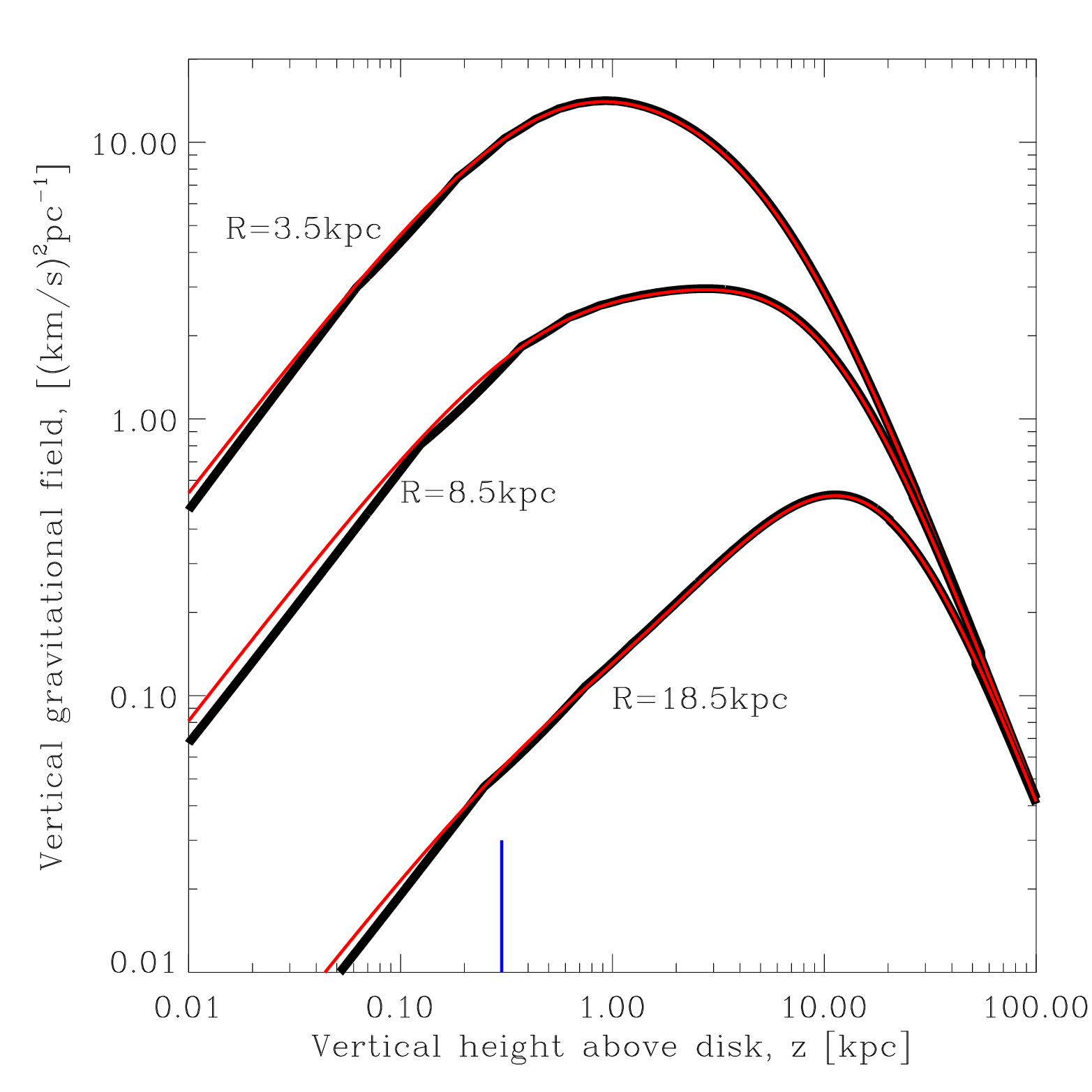}
}
\subfigure{
\includegraphics[angle=0,width=8.0cm]{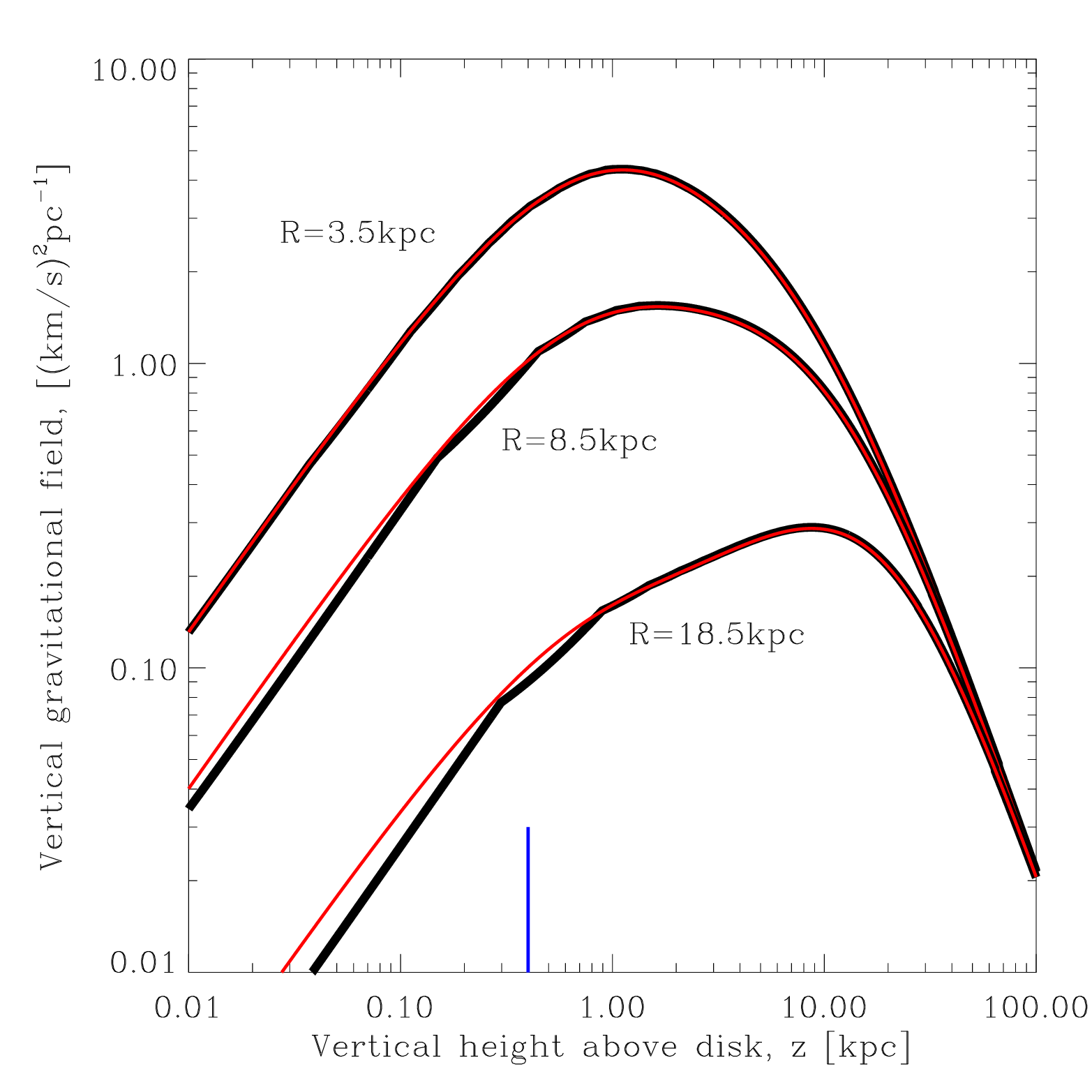}
}
\\
\subfigure{
\includegraphics[angle=0,width=8.0cm]{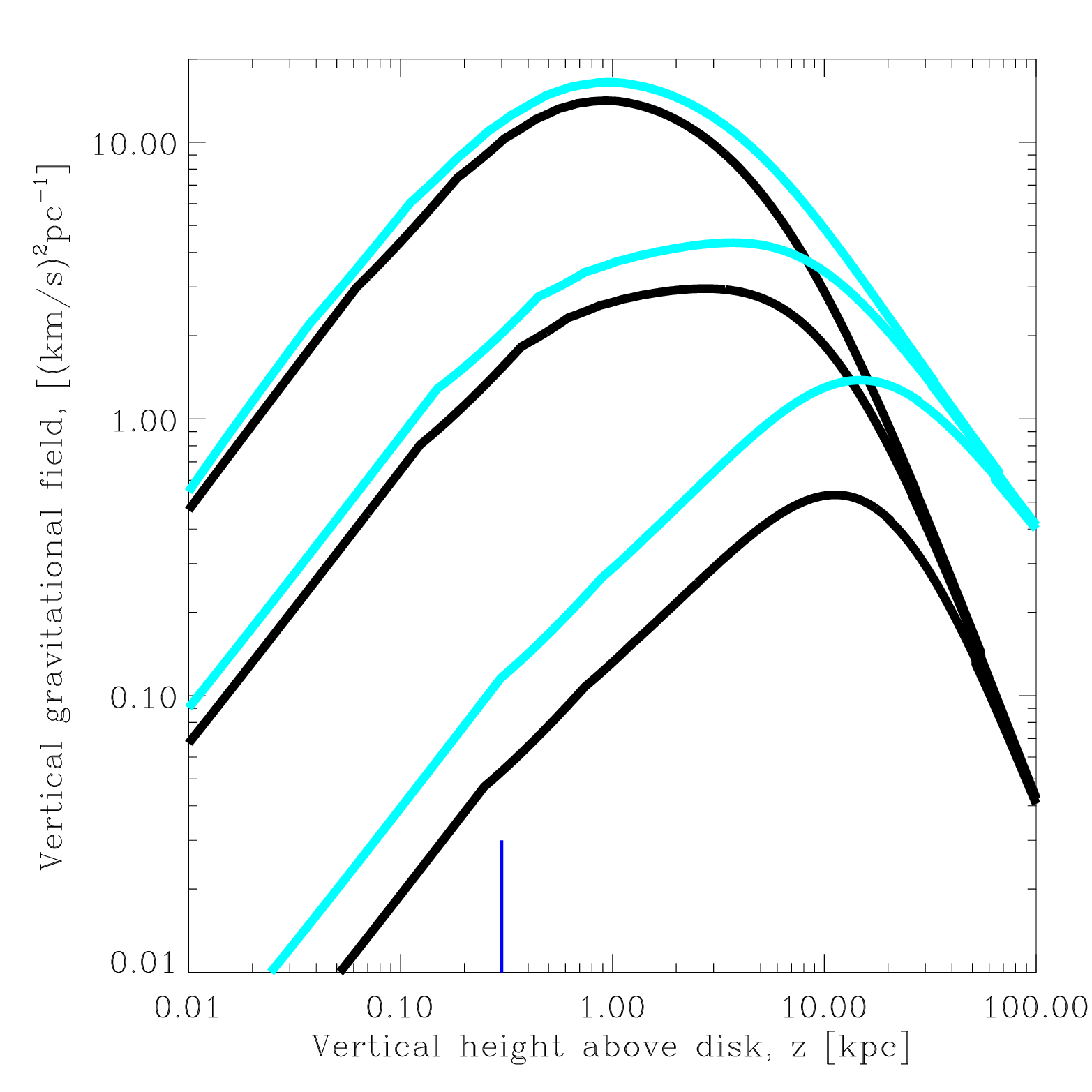}
}
\subfigure{
\includegraphics[angle=0,width=8.0cm]{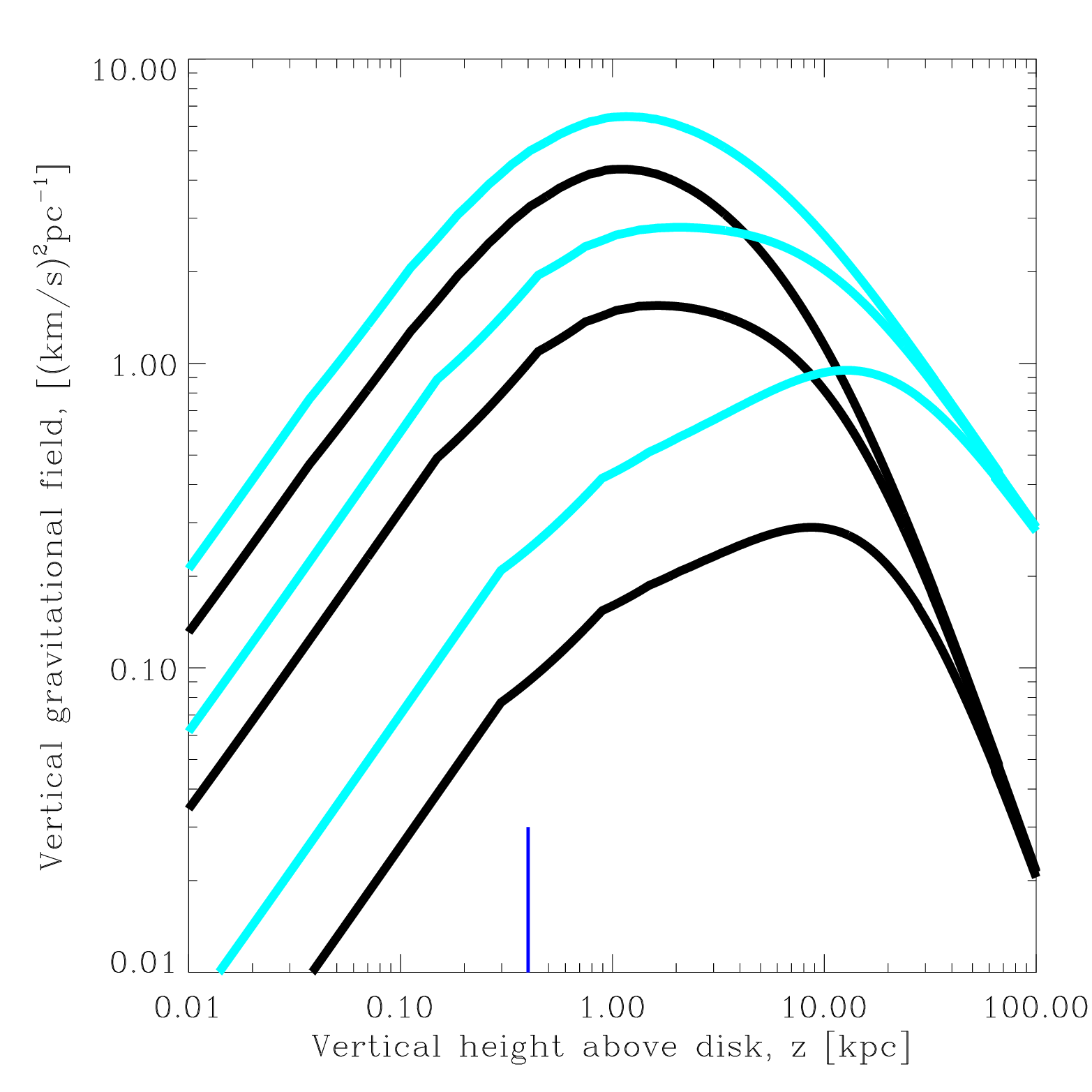}
}
\end{tabular}
\caption{ Vertical gravity profile through three discrete cylindrical radii in the disk mid-plane $R=3.5$, 8.5 and 18.5~kpc. (top panels) Comparison of the Newtonian theoretical vertical gravity profile against the computations from a Poisson solver for two different model galaxies. The galaxies are composed, as per Eq~\ref{eqn:jdisk}, of a radial exponential disk and exponential scale-height. The left hand panels use a galaxy with scale-length $h_R=2.5$~kpc, scale-height $h_z=0.3$~kpc and total mass $M_d=10^{11}\msun$. The right hand panels use a galaxy with scale-length $h_R=4$~kpc, scale-height $h_z=0.4$~kpc and total mass $M_d=5\times10^{10}\msun$. The computations from the Poisson solver with Newtonian dynamics use the thick black line and the theoretical values found using Kuijken \& Gilmore (1989) Eq~27 use the thin red line. The short blue vertical lines mark the scale-height of the model disk. The match is generally very good for $z>0.1kpc$.
(bottom panels)
Comparison of the MOND (using $\gamma=1$; turquoise lines) and Newtonian (black lines) vertical gravity profiles as calculated by the Poisson solver. The galaxies used in both panels are the same as the ones used in the top panels. In these units, the MOND acceleration parameter is $a_0=3.6~(\kms)^2\rm pc^{-1}$. Clearly the MOND boost to the gravity is relatively less significant for the higher surface density galaxy.}
\label{fig:grav}
\end{figure*}

\section{Preliminary Modelling}
\subsection{Bulge-Disk decomposition}
\protect\label{sec:bdd}
The inclination corrected surface brightness profiles presented by DMSvi were analysed with a Markov Chain Monte Carlo (MCMC) approach to fit bulge plus disk surface brightness models using Eqs~\ref{eqn:sersic} and \ref{eqn:sigdisk}. The seeing was simultaneously accounted for with a Gaussian convolution of appropriate radius (see tables 3 and 4 of DMSvii). All 5 surface brightness parameters (2 disk and 3 bulge) were fitted, but the bulge surface brightness parameters were found to be degenerate with each other due to seeing effects and the lack of data points at low radii. Thus, the bulge parameters were fixed at the maximum likelihood values, which allowed better sampling of the more important disk parameters. In Figs~\ref{fig:like0}-\ref{fig:like2} the posterior probability distributions are presented for the two disk surface brightness parameters of each galaxy. For the disk scale-length, the range found by DMSvi (red curve) is also plotted.

The best fits to the surface brightness profiles are given in Figs~\ref{fig:1like0}-\ref{fig:1like2} and are generally good. Both linear radius and log radius are plotted on the $x$-axis to expose the quality of the fits to the bulge and the outer disk. The fitted scale-lengths found here (see Figs~\ref{fig:like0}-\ref{fig:like2}) are generally consistent with those fitted by the DMS, although theirs have smaller error ranges. 

Once the bulge and disk surface brightness profiles were established, an Abel transform was used to de-project the bulge surface brightness, which was stored numerically. The disk has cylindrical symmetry, so it does not require de-projection. Following this, rejection sampling was used to generate N-particle representations of the galaxies, with half of all particles representing the stellar disk and the other half split evenly between the stellar bulge and the atomic and molecular gas disks. The fitted scale-lengths are used for all the following analysis, but they are not used to generate the N-particle representations of the stellar disk. Instead, the ``observed'' surface brightness of the disk is sampled after subtracting the fitted bulge - since the fit to the surface brightness can be poor at large radii.

In this analysis the static N-particle representations are only used to find the potential and gravitational field, there is no N-body evolution of the simulated galaxies. The masses recovered after generating the N-particle representations of each component of each galaxy generally agreed very well with the masses reported by the DMS.

\subsection{The MOND Poisson solver}

The MOND Poisson solver described in \cite{angus12} is used to compute the radial and vertical gravitational fields of the N-particle galaxy models. The code solves the modified Poisson equation of the Quasi-linear version of MOND (QUMOND) given by Eq~\ref{eqn:mondpois}. The code uses a 3D grid and the cloud-in-cell technique to numerically discretise the 3D density of an N-particle distribution. It then employs finite differencing and multigrid methods to iterate from a test potential to the final potential which accurately reflects the density.

\subsubsection{Comparison of theoretical and numerical vertical gravity}
\protect\label{sec:galmods}
The Poisson solver is used to compute the gravitational field of the baryons in the radial direction, ${d\Phi_{bar} \over dR}(R)$, as a function of radius, to find the model rotation speed (Eq~\ref{eqn:vrot}). Simultaneously, the gravitational field of the baryons is solved for in the vertical direction, $d\Phi_{bar}(R,z) \over dz$, as a function of height above the disk, at several discrete radii: $R=0.5$ to 9.5~kpc in steps of 1~kpc and then $R=11.5$, 14.5, 18.5 and 25.5~kpc. In MOND, both gravitational fields noted above are equivalent to the gradient of the total potential (i.e. $\Phi_{bar}\equiv \Phi_{tot}$) used in Eqs~\ref{eqn:vrot} and \ref{eqn:rho*} and thus the rotation speed can be straight-forwardly calculated and  Eqs~\ref{eqn:rho*} and \ref{eqn:sig*} can be integrated to find $\sigma_z$ at different radii, $R$.

The most important, and difficult to compute, quantity is the vertical gravity profile at large radii. For an isolated, double exponential disk, like that introduced in Eq~\ref{eqn:jdisk}, the Newtonian vertical gravity profile at a given radius can be calculated numerically - as described in detail by \cite{kuijken89}.

In the top panel of Fig~\ref{fig:grav} the Newtonian vertical gravity profile is plotted for three discrete cylindrical radii $R=3.5$, 8.5 and 18.5~kpc. The left and right hand panels correspond to two different double exponential disk galaxies (as per Eq~\ref{eqn:jdisk}). The left hand panel has scale-length $h_R=2.5$~kpc, scale-height $h_z=0.3$~kpc and total mass $M_d=10^{11}\msun$. The right hand panel has scale-length $h_R=4$~kpc, scale-height $h_z=0.4$~kpc and total mass $M_d=5\times10^{10}\msun$. The computation from the Poisson solver using Newtonian dynamics is the thick black line and the theoretical value using Eq~27 of \cite{kuijken89} is the red line. The agreement is generally very good, except for small heights above the disk. This is due to a combination of the limit of spatial resolution at large radii, owing to the centrally refining mesh, and limited particles at large radii. The average percentage error for these discrete radii (3.5, 8.5, 11.5, 14.5, 18.5 and 25.5~kpc) between the numerical calculation of $\sigma_z$ (using Eq~\ref{eqn:sigred}) and the theoretical values are 0.05, 0.8, 0.7, 5.8, 5.2 and 4.4\%. Given that only 6 of the 30 galaxies have vertical velocity dispersion data points beyond 11~kpc and the errors on those data points are typically more than 10\%, this is more than adequate.

In the bottom two panels of Fig~\ref{fig:grav} the Newtonian (black lines) and MOND ($\gamma=1$; turquoise lines) vertical gravity profiles are compared, both of which were computed with the Poisson solver. The MOND profiles are strongly boosted relative to the Newtonian profiles, however, the higher surface density galaxy (left hand panel) receives less of a boost than the other.

\subsection{Error budget}
\protect\label{sec:errors}

The extra information that allows the DMS to close their set of equations is that, in general, disk scale-heights are observed to correlate with their scale-lengths. These scale-heights and scale-lengths have been fitted to edge-on galaxies and certain assumptions are made in their modelling (like constant inclination, that the disks are well described by exponential or $sech^2$ distributions). The DMS made an analysis of literature measurements and derived a simple relationship between $h_R$ and $h_z$ such that in units of kpc (see DMSii)
\beq
\protect\label{eqn:kreg}
h_z \sim 0.2h_R^{0.633}.
\eeq
This relation has a 1$\sigma$ scatter of roughly 25\% (DMSii). 

In the analysis presented here, the scale-height and disk $M/L$ are fitted to the observed vertical velocity dispersions and rotation curves of the sample of 30 galaxies. After the distribution of fitted scale-height and scale-lengths is known, they will be compared for consistency with the direct observations of $h_R$ and $h_z$ for the sample of edge on galaxies compiled by the DMS. Therefore, a simultaneous fit must be made to the observed vertical velocity dispersions and rotation curves. A critical concern is the contribution of each of the two datasets to the overall likelihood. Since the vertical velocity dispersions are the primary data set, the errors on each data point as taken as computed by the DMS. Given that there is the potential for some systematic errors from disk warping, the rotation curve should not be given too large a weighting, especially the inner parts. The decision was made to increase the error bars on the rotation curve data points to 10~$\kms$. Lastly, the two separate reduced $\chi^2$s from fitting the vertical velocity dispersion profile and the rotation curve are combined.

Therefore, the likelihood is given by

\bey
\protect\label{eqn:like}
\nonumber -2log_e\mathcal{L}&=&n_{RC}^{-1}\Sigma_{i=1}^{n_{RC}}\left({V_{mod}(R_i)-V_{obs}(R_i) \over 10 km/s }\right)^2\\
&+&n_{VVD}^{-1}\Sigma_{j=1}^{n_{VVD}}\left({\sigma_{z,mod}(R_j)-\sigma_{z,obs}(R_j) \over \sigma_{z,obs,error}(R_j) }\right)^2,
\eey

where $n_{RC}$ and $n_{VVD}$ are the number of relevant data points in the rotation curve and vertical velocity dispersion profile respectively. The prior then multiplies $\mathcal{L}$ to give the un-normalised posterior probability.

\section{Primary results}
\protect\label{sec:primary}
\subsection{Parameter posterior probabilities}
\protect\label{sec:parlike}

MCMC sampling was used to find the posterior probability distribution for each of the free parameters: disk scale-height ($h_z$), disk $M/L$ ($\Upsilon_d$) and inclination ($i$). The scale-height was varied by considering the ratio of the fitted scale-height to the one derived from observations (Eq~\ref{eqn:kreg}). A broad Gaussian prior of width 1.5 was placed on this ratio. The prior placed on the disk $M/L$ was centred on $\Upsilon_d=0.3~\msun/\lsun$ and had a width 0.5~$\msun/\lsun$. The ratio between the fitted inclination and the luminous TF relation inclination (from Eq~\ref{eqn:tfr}) received a fairly tight Gaussian prior of 0.15 (or 15\%), since there should be little deviation from the TF relation. The main reason for including inclination as a free parameter is just in case there is a sharp increase in posterior probability for a small change in inclination. The full expression for the likelihood is given by Eq~\ref{eqn:like}. This defines how the goodness of fit for the two data sets is combined: rotation curve and vertical velocity dispersion.

The re-normalised posterior probability distributions of each of the three parameters (scale-height, inclination and $M/L$) is plotted in Figs~\ref{fig:2like0}-\ref{fig:2like2} for each galaxy individually. There are two main lines in each of the three columns: MOND with $\gamma=1$ and 2 (see Eq~\ref{eqn:nualp}) black lines, solid and dashed respectively. The second column (inclination) also has the curve of the prior (turquoise). The third column (M/L) has a green line which represents the Newtonian gravity (with DM halo) $M/L$ found by the DMS (DMSvi), shown only for reference. All the lines are re-normalised to have the same maximum posterior probability. The two MOND fits vary little in terms of goodness of fit.

From the left hand column it is clear that the majority of the fits require substantially lower scale-heights than those derived from observations (Eq~\ref{eqn:kreg}), which in those panels are unity. There is a preference with most galaxies to have a higher inclination because that decreases the amplitude of the rotation curve. A lower rotation curve requires a lower $M/L$. A lower $M/L$ allows a larger scale-height - which allows better agreement with the observations of edge-on galaxies (Eq~\ref{eqn:kreg}). However, the fairly tight prior on the inclination prevents it from changing substantially. Usually it changes by less than 10\%.

In Table~\ref{tab:sims} various 1$\sigma$ confidence ranges for the fitted scale-heights and stellar mass-to-light ratios are presented using MOND with varying constraints (from interpolating functions and which parameters are left free). The preferred scale-heights, those from Eq~\ref{eqn:kreg} and used by the DMS, are given for each galaxy (column 2). For reference (column 3) the $M/L$ found by the DMS when using Newtonian gravity and DM halos is shown, followed by the best fit MOND scale-heights (as a ratio between the fitted scale-height and the one derived from observations of edge-on galaxies - Eq~\ref{eqn:kreg}) and the $M/L$ for three different interpolating functions (the two used above and one other). These fits are for the scenario where the scale-height and $M/L$ are barely constrained, but the inclination is tightly constrained by its prior.

\subsubsection{Comparison of the MOND fits with the DMS data}
In Figs~\ref{fig:3like0}-\ref{fig:3like2} the best fits to the vertical velocity dispersion (first and third columns) and rotation curve (second and fourth columns) are plotted in the MOND case with $\gamma=1$. The two interpolating functions are not plotted together because different best fit inclinations mean the data points vary in each case. There is little difference between the two MOND cases in terms of quality of fit. Most fits to the vertical velocity dispersion are good, but the quality of fits to the rotation curves range from poor to good. Rotation curves with good fits include the galaxies UGC 448, 1081, 1087, 1908, 3091, 4036, 4368, 4380, 7244, 6918, 8196, 9965, 11318, 12391. Quite good fits are found for UGC 463, 1529, 1635, 1862, 3701, 4256; whereas
the fits to UGC 3140, 3997, 4107, 4458, 4555, 4622, 6903, 7917, 9177, 9837 are poor.

MOND fits to the rotation curves of nearly face-on galaxies are very sensitive to the inclination (see e.g. \citealt{deblok98}). In the outer parts, they depend on the fourth power of $1/sin(i)$.  Warping of the disk might be invoked to improve certain rotation curve fits, however, in some cases the DMS galaxies have already been corrected for warps.

To compare with models that use the scale-height from Eq~\ref{eqn:kreg}, the vertical velocity dispersion (red line) is over-plotted for each galaxy using that larger scale-height, but the $M/L$ and inclination of the best fit are kept. In many cases, the red line has a far greater amplitude than the data points and best fit (black line). There are, however, a few cases where the red and black lines are close together, such as UGC 1862, UGC 4368 and UGC 4458 - although the latter only has a single relevant data point.

\subsection{Comparison of the fitted $h_z$ vs $h_R$ with observations}

From fitting the observed surface brightness profiles of the DMS galaxies, a value for the radial scale-length was estimated for each galaxy disk. As can be seen in Figs~\ref{fig:like0}-\ref{fig:like2} (second and fourth columns) the radial scale-length fitted by DMSvi is generally slightly larger than the best fit found here, but the majority are consistent within the errors. This fact makes the analysis presented here slightly more favourable to MOND. The difference in fitted scale-lengths is in part due to the fact that in DMSvi a finite central region was used to fit the scale-length, whereas the full extent of the galaxy is used here. Thus, for any galaxies whose disk surface brightness is not well fit by a single exponential there is a difference. For example, UGC 4036 has a shallow inner surface brightness profile and then a steep outer decline, thus the scale-length found here naturally has a smaller scale-length. Recall that the N-particle realisations (\S\ref{sec:bdd}) use the observed surface brightness, not the fitted exponential disk.

\begin{figure*}
\includegraphics[angle=0,width=17.0cm]{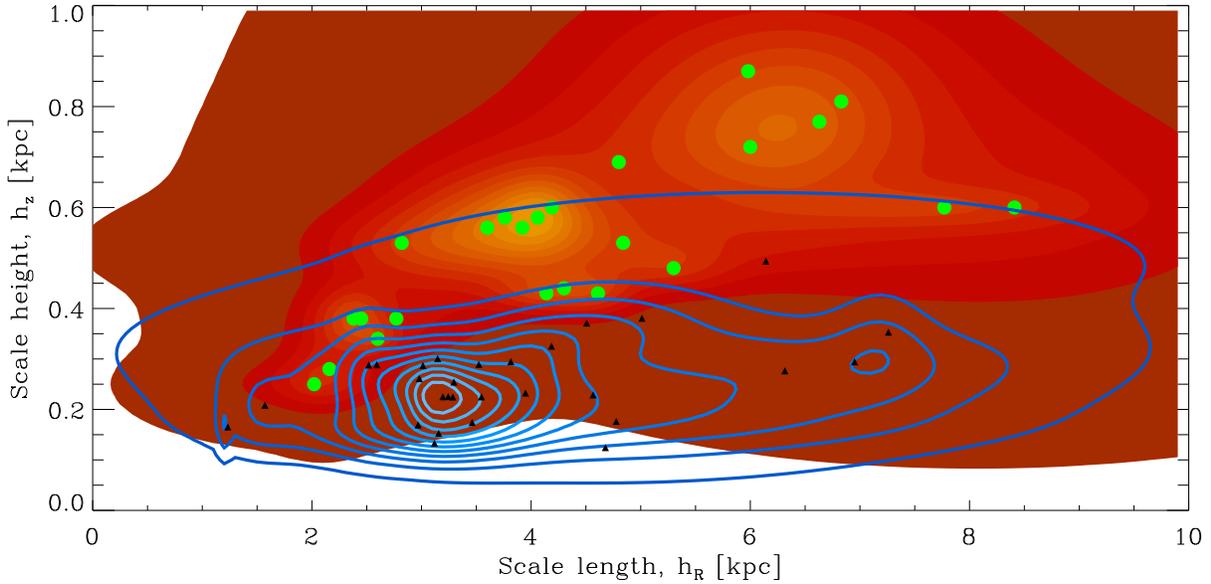}
\caption{Contours of the posterior probability distribution of galaxies in the $h_z$ vs $h_R$ plane. The blue contours and the black triangles correspond to the scale-lengths fitted to the surface brightness profiles of the DMS galaxies and the scale-heights found from the MOND ($\gamma=1$, $a_0=3.6(\kms)^2\rm pc^{-1}$) fits to their vertical velocity dispersions. The measurements of scale-lengths and scale heights from observations of edge-on galaxies by K02 have red shaded contours with green circles. The re-normalised contour levels are 1.0, 0.91, 0.82, 0.73, 0.64, 0.55, 0.46, 0.37, 0.28, 0.19, 0.1, 0.0001. The two sets of contours are highly inclined relative to each other.}
\label{fig:simp}
\end{figure*}

Similarly, by fitting the vertical velocity dispersion profile and the rotation curve of each DMS galaxy, a confidence range for the scale-height of each galaxy has been derived. In principle, this combination of $h_z$ and $h_R$ parameters can be compared to the measurements of scale-lengths and scale-heights of a sample of edge-on galaxies. In Fig~\ref{fig:simp} the $h_z$ vs $h_R$ diagram for the MOND fits ($\gamma=1$, $a_0\sim3.6(\kms)^2\rm pc^{-1}$) is plotted along with the observations of these parameters for edge-on galaxies made by K02.

In order to visualise this more plainly, contours in the $h_z$ vs $h_R$ plane were added to Fig~\ref{fig:simp}. Here, the re-normalised posterior probability density of all individual data points are co-added on a fine grid, with each galaxy receiving equal weighting. A point with larger error bars will spread its probability density over a larger range. The blue contours represent the co-added MOND probability densities and the black triangles give the location of each individual galaxy point. The error bars are not included to minimise clutter. The measurements by K02 have red shaded contours with green circles. The MOND contours are significantly inclined relative to the K02 contours. The equivalent plot for the $\gamma=2$ interpolating function is almost identical.

%\begin{figure*}
%\includegraphics[angle=0,width=17.0cm]{stan.ps}
%\caption{As per Fig~\ref{fig:simp}, but using $\gamma=2$.}
%\label{fig:stan}
%\end{figure*}

\subsection{Two dimensional Kolmogorov-Smirnov test}
\protect\label{sec:2dks}

To demonstrate statistically the the large offset in the $h_z$ vs $h_R$ plane between the K02 data points in Fig~\ref{fig:simp} and the fitted MOND points, a 2D Kolmogorov-Smirnov (KS) test (see \citealt{press92}) chapter 14.7) was employed. In this test, the first K02 data point was taken and made the origin of four quadrants. Then the fraction of K02 points in each quadrant and the fraction of MOND points in the same quadrants. The largest difference between the two fractions in a single quadrant is stored and then the rest of the K02 points are cycled through, making each one the origin in turn. The largest of the largest differences is stored and then the procedure is repeated centring on the MOND points. The largest difference in fractions is ordinarily the same regardless of whether the K02 or MOND points are central. This number is the statistic of merit for the 2D KS test and provides a significance level that the two samples originate from the same parent population. 

Since the data points have fairly large error bars, if the mean of a particular data point lies in the first quadrant, it is not ideal to add the whole point to that quadrant. Instead, the fraction of the point inside each quadrant determined by the point's error bars is added.

According to the 2D KS test, the significance levels that the MOND fits with $\gamma=1$ and 2 come from the same parent distribution as the K02 sample are $4.4\times10^{-5}$ and $1.4\times10^{-4}$ respectively. Using the interpolating function $\delta=4$ the significance level is $1.0\times10^{-4}$. Comparing with DM models, it is easy to obtain significance levels of order unity, so the method seems robust.

As a separate test, a straight line was fitted to each data set of $h_z$ vs $h_R$ with a fixed intercept. The resulting gradients differ at the 2$\sigma$ level, regardless of interpolating function. Using a variable intercept only improves the agreement slightly.

\subsection{Superposition of $M/L$ distributions}
\begin{figure}
\includegraphics[angle=0,width=8.5cm]{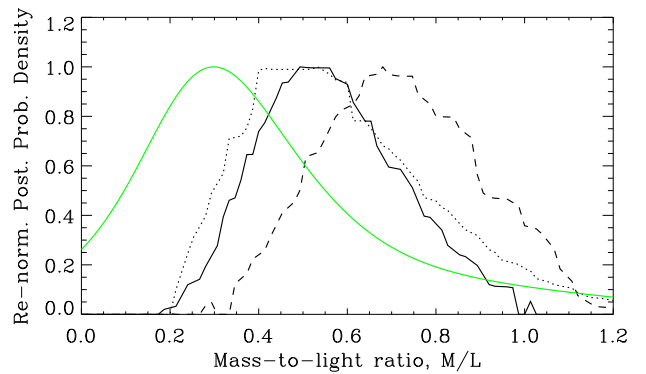}
\caption{Co-addition of all 30 re-normalised posterior probability profiles for the fitted $M/L$ parameter. MOND fits for the interpolating functions using Eq~\ref{eqn:nualp} with $\gamma=1$ and 2 are represented by black solid and dashed curves respectively and $\delta=4$ (from Eq~\ref{eqn:nudel}) has a dotted curve. The co-added $M/L$ found by the DiskMass Survey in DMSvi has a green curve.}
\label{fig:cumfits}
\end{figure}

\protect\label{sec:superpos}
In the approach presented here to modelling the DMS data the $M/L$ is fitted for as well as the scale-height. At near infrared wavelengths a smaller variation of the $M/L$ is expected from one galaxy to the next than at optical wavelengths (\citealt{belldejong,bell07}). To find the posterior probability distribution of $M/L$ across the full sample of galaxies, the summation of the re-normalised posterior probability density for the $M/L$ of all 30 galaxies is plotted in Fig~\ref{fig:cumfits}. This is done for three MOND interpolating functions and the DMS computation of $M/L$ in the DM scenario. This means the same weighting is given to each galaxy's $M/L$ posterior probability density in the summation. Recall that in the fitting process a Gaussian prior is used with width 0.5$\msun/\lsun$ centred on $M/L=0.3\msun/\lsun$.

The $\gamma=1$ MOND interpolating function (solid black line) has an average best fit $M/L \simeq 0.55 \pm 0.15$, whereas the $\gamma=2$ MOND interpolating function (dashed black line) requires significantly larger values, $M/L \simeq 0.7 \pm 0.2$. Bear in mind that the $\gamma=2$ produces less of a boost to the gravity than the $\gamma=1$ interpolating function at intermediate gravities around $\sim a_0$. The $\delta=4$ interpolating function (which uses Eq~\ref{eqn:nudel}) has a similar $M/L$ distribution as $\gamma=1$. There is no impact on the derived $M/L$ values from the weak prior. The green line, given only for reference, is the $M/L$ derived from the DMS analysis with Newtonian gravity and DM.

\subsection{Why does MOND need such thin disks?}
The results presented here suggest that MOND requires disks that have roughly half the vertical scale-height as those inferred from observations of edge on galaxies (DMS II or Eq~\ref{eqn:kreg}). There are two key differences between the MOND vertical velocity dispersions and those using DM. The first effect is related to the required $M/L$ of the galaxies. Comparing a disk with the same surface brightness in MOND and Newtonian gravity, the Newtonian disk is surrounded by a dark matter halo. This dark matter halo can account for the radial gravitation field that produces the observed rotation speed. This means the $M/L$ of the Newtonian disk is only weakly constrained by the rotation curve (\citealt{vanalbada85}). In MOND, the $M/L$ is essentially fixed by the rotation speed because it is the only parameter that can be varied to allow a match between the observed and expected rotation speeds. Based on the results presented here, the required $M/L$ of the disk is a factor of between 1.5 and 2.5  higher in MOND (cf. Fig~\ref{fig:cumfits}), depending on the interpolating function used. Since the $M/L$ is typically much larger in MOND, the resulting vertical gravitational field is larger than the dark matter equivalent. Moreover, even if the disks in MOND and Newtonian gravity had the same $M/L$, the MOND gravitational field is still boosted relative to the Newtonian one as seen in Fig~\ref{fig:grav}. This follows because the nearly spherical dark matter halo typically has a weak influence on the vertical gravitational field within the disk.

The influence of MOND on the vertical gravitational field, relative to Newton, for the same disk is of course greater when the surface density is low. This is demonstrated by comparing the vertical gravitational field of a high surface brightness galaxy (Fig~\ref{fig:grav} bottom left panel) with a lower surface brightness one (Fig~\ref{fig:grav} bottom right panel).

In relation to the equations of \S\ref{sec:dyn}, it can be seen that increasing $\Upsilon_d$ or $h_z$ increases the amplitude of the model $\sigma_z$. However, in MOND only increasing $\Upsilon_d$ increases the model rotation curve at intermediate to large radii. Thus, once $\Upsilon_d$ is fixed by the rotation curve, only $h_z$ can be varied to fit the observed vertical velocity dispersion. Since the observed vertical velocity dispersions are typically much lower than predicted by MOND, $h_z$ must be decreased to fit them. It is for these reasons, the fitted scale-heights in MOND are significantly lower than those expected from observations (Eq~\ref{eqn:kreg}).

\subsection{Central dynamical surface density versus central surface luminosity density}
\cite{swaters14} used the DMS data to present a correlation between the extrapolated disk central surface luminosity density, from the photometry, and the extrapolated disk dynamical central surface mass density (see \S\ref{sec:dmsmethod}). The majority of galaxies are consistent with the relation $\Sigma_{dyn}(0)=0.3\msun/\lsun I_*(0)$, but at the low surface luminosity density end a clump of 6 galaxies lie above the relation. In the Newtonian context, this can be attributed to greater relative DM dominance for low surface luminosity density galaxies (\citealt{swaters14}), but this is also reminiscent of the MOND transition in rotation curves of spiral galaxies, where the effect becomes pronounced at low surface densities. Thus, a prediction for dynamical surface density versus central surface luminosity density in MOND would be valuable.

\begin{figure}
\includegraphics[angle=0,width=8.50cm]{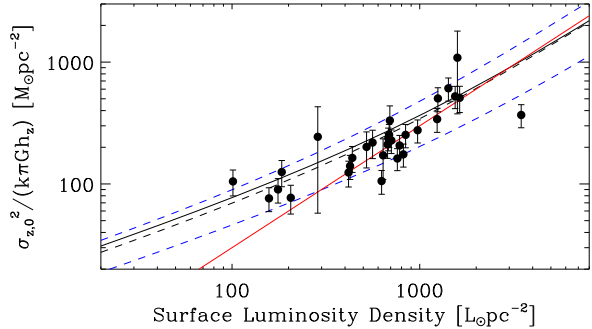}
\caption{Disk central dynamical surface density versus disk central surface luminosity density relation. The data points correspond to each galaxy in the DiskMass Survey and are reproduced from Swaters et al. (2014). The red line illustrates $\Sigma_{dyn}(0)=0.3\msun/\lsun I_*(0)$. The other lines show MOND predictions: the dashed lines use scale-heights that are twice as thin as those derived from fits to observations of edge-on galaxies (Eq~\ref{eqn:kreg}), and the solid black line uses the scale-heights derived from fits to observations of edge-on galaxies. The black lines all use $M/L=0.6$. The upper blue dashed line uses $M/L=0.9$ and the lower one uses $M/L=0.3$. The apparent flattening of the central velocity dispersions at low surface brightnesses is consistent with MOND.}
\label{fig:dmdvar}
\end{figure}

To this end, a representative galaxy from the DMS was selected - UGC 4036. The stellar bulge and all gas was removed from the N-particle realisation, thus it has the same stellar disk surface brightness profile as UGC 4036, and the same vertical profile. For this model, the central vertical velocity dispersion, $\sigma_z(0)$ from Eq~\ref{eqn:sigred}, can be found for a chosen $M/L$. From here, the surface brightness of the stellar disk can be scaled to compute $\sigma_z(0)$ at a broad range of central surface luminosity densities which allows a comparison with the data from \cite{swaters14}, which is done in Fig~\ref{fig:dmdvar}.

The data points come from \cite{swaters14} Fig~1 (left hand panel) and both coordinates are extrapolated from exponential fits to the observed projected disk surface luminosity density and observed vertical velocity dispersion. The red line is the relation $\Sigma_{dyn}(0)=0.3\msun/\lsun I_*(0)$. The black solid line is the MOND prediction with $M/L=0.6$ and the scale-height expected from observations of edge-on galaxies (DMSii; Eq~\ref{eqn:kreg} in this paper). The black dashed line has the same $M/L$, but used disks that are twice as thin, which MOND has been shown here to require in order to mesh with the DMS data. The two blue dashed lines also use disks twice as thin as observations of edge-on galaxies, with the upper line using $M/L=0.9$ and the lower line using $M/L=0.3$. Thus, MOND provides a natural explanation for the deviation of low surface luminosity density galaxies from the central dynamical surface density versus central surface luminosity density relation.

\section{Secondary results: Varying additional parameters}

In this section the impact of other variables on the results from \S\ref{sec:primary} is considered.

\subsection{Variation of the Stellar Velocity dispersion Ellipsoid}
\protect\label{sec:vsve}

\begin{figure}
\includegraphics[angle=0,width=8.5cm]{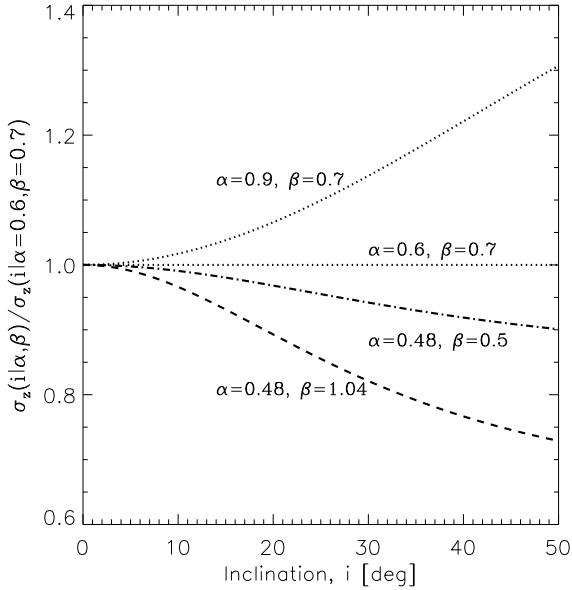}
\caption{Ratio between the $\sigma_z$ found from Eq~\ref{eqn:vellip}, using certain $\alpha$ and $\beta$ parameter combinations, and the default values used by the DMS (which are $\alpha=0.6$ and $\beta=0.7$ as per DMSvi) against inclination. The combination of $\alpha$ and $\beta$ used for each line is made clear in the figure. Varying $\alpha$ can have a significant impact on the derived velocity dispersion for increasingly inclined galaxies.}
\label{fig:vvdvar}
\end{figure}

As discussed in \S\ref{sec:svvd}, the stellar velocity dispersion ellipsoid (SVE) parameters are fixed to those used by the DMS i.e. $\alpha=0.6$ and $\beta=0.7$. To demonstrate the degree to which a reasonable variation in these parameters might affect the results, in Fig~\ref{fig:vvdvar} the vertical velocity dispersion ratio between three combinations of parameters and the default values are plotted against inclination.

Using different values for $\alpha$ in Eq~\ref{eqn:vellip} can make a sizeable difference to the derived $\sigma_z$, especially for large inclinations. The dotted line of Fig~\ref{fig:vvdvar} shows that increasing $\alpha$ to 0.9, which is 2$\sigma$ above the default value used by DMSvi, would increase the derived $\sigma_z$ by around 10\% for standard DMS inclinations of $25^{\circ}$. On the other hand, decreasing $\alpha$ to 0.48 and using $\beta=1.04$ (dot-dashed line in Fig~\ref{fig:vvdvar}), which was the value derived by \cite{westfall11} for UGC~463, would {\it decrease} the derived $\sigma_z$ by around 15\% for $i=25^{\circ}$.

To test the impact of these SVE parameters on the fits to the vertical velocity dispersions and rotation curves, MCMC modelling was used to simultaneously fit $M/L$, $\alpha$ and $\beta$, with inclination fixed by the TF relation and $h_R/h_z$ set by observations of edge-on galaxies (Eq~\ref{eqn:kreg}). Here, only flat priors are used such that $0<\alpha<1$ and $0<\beta<2.5$. The $M/L$ has a Gaussian prior of width 0.5 centred on 0.3. The most relevant parameter is $\alpha$, due to its impact on Eq~\ref{eqn:vellip}.

\begin{figure}
\includegraphics[angle=0,width=8.50cm]{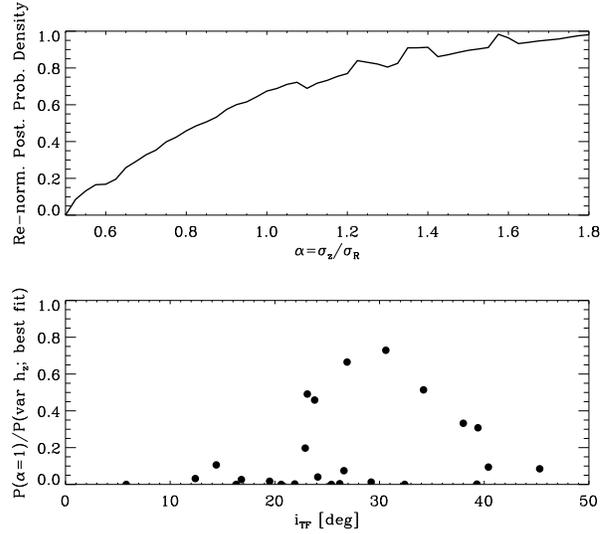}
\caption{(top panel) Superposition (for all 30 galaxies) of the re-normalised posterior probability density as a function of $\alpha$. (bottom panel) Ratio between the posterior probability of the MOND fit to the DMS data with fixed scale-height and inclination, but $\alpha=1$ (along with best fit $\beta$ and $M/L$), to the benchmark fit. This is plotted against the inclination of the galaxy derived from the luminous TF relation. The MOND parameters $\gamma=1$ and $a_0=3.6~(\kms)^2\rm pc^{-1}$ are used.}
\label{fig:alpbet}
\end{figure}

In the top panel of Fig~\ref{fig:alpbet} the co-addition (for all 30 galaxies) of the re-normalised posterior probability is plotted as a function of $\alpha$. Each of the 30 galaxies are given equal weighting, regardless of their individual posterior probability. This shows the impact of increasing $\alpha$ from the nominal value of 0.6 used here and by the DMS. The posterior probability increases roughly a factor of two and a half from $\alpha=0.6$ to 1. 

This increase does not necessarily mean that the fit qualities are good. To demonstrate this, it is worth comparing the fits where $\alpha$ is a free parameter (but scale-height is fixed) to the fits where the scale-height is free. These fits where scale-height is free represent, for the majority of galaxies, a good quality fit to the vertical velocity dispersion. For the remainder of this section, this fit is referred to as the {\it benchmark} fit. $\alpha$ does not influence the fit to the rotation curve.

In the bottom panel of Fig~\ref{fig:alpbet} the $y$-axis is the ratio between the posterior probability of the fit with fixed scale-height (Eq~\ref{eqn:kreg}) and inclination (luminous TF relation), $\alpha=1$ and best fit $\beta$ and $M/L$ to the benchmark fit. This is plotted against the inclination of the galaxy derived from the luminous TF relation to show that increasing $\alpha$ may help with some (but not all) of the high inclination galaxies, but the majority of the sample would have extremely poor fits to the DMS data even with $\alpha=1$.

\subsection{Fixed scale-height, unconstrained inclination}
\begin{figure}
\includegraphics[angle=0,width=8.5cm]{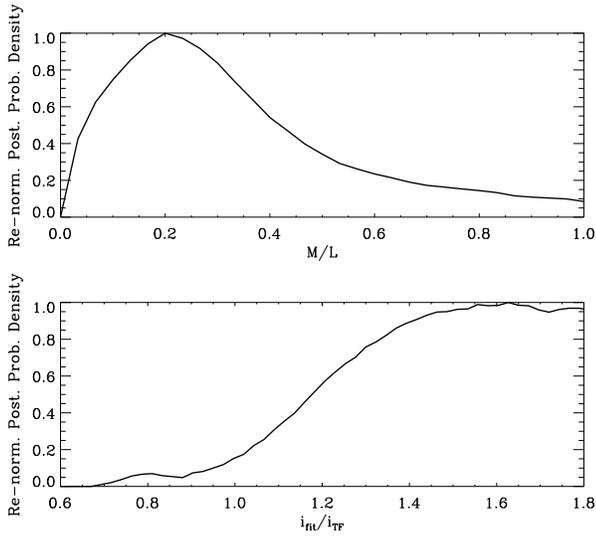}
\caption{Impact of varying inclination on the goodness of fit and the $M/L$. These models use fixed scale-heights (Eq~\ref{eqn:kreg}), but the inclination is left completely unconstrained. (top panel) Co-added posterior probability density as a function of $M/L$, for all 30 galaxies, required for high quality fits to the DMS data when the inclination is left free. (bottom panel) Co-added posterior probability density against the ratio of the fitted inclination to the inclination derived from the luminous TF relation (V01). The inclinations required for good fits to the DMS data are far larger than the inclinations derived from the luminous TF relation.}
\label{fig:fixhz}
\end{figure}

A relevant question is how far the fitted inclination would have to deviate from the luminous TF relation before a good fit to the DMS data could be achieved, whilst remaining consistent with the $h_z$ values derived from fits to the photometry of edge-on galaxies. To do this, the scale-height was fixed to the aforementioned value (Eq~\ref{eqn:kreg}) for each galaxy and fitted only for the inclination and $M/L$. The parameters $\gamma=2$ and $a_0=3.6~(\kms)^2\rm pc^{-1}$ were used. For the sake of brevity, in Fig~\ref{fig:fixhz} only the co-added, posterior probability densities for the two parameters are plotted.

The top panel shows the $M/L$, which is shifted to values roughly three times smaller (cf. dashed line of Fig~\ref{fig:cumfits}) than the original models with scale-height free. In the last column of Table~\ref{tab:sims} the $M/L$ range is given for each galaxy in this scenario individually. The $M/L$ values are much smaller than the other models where inclination is constrained. In the bottom panel of Fig~\ref{fig:fixhz} the inclination, which is given as a ratio of the fitted value ($i_{fit}$) to the luminous TF relation value ($i_{TF}$), is plotted. With an unchanged inclination ($i_{fit}=i_{TF}$), the quality of the fits to the majority of the DMS galaxies are poor, but by increasing the fitted inclination to $i_{fit}/i_{TF}=1.4$ the fits are of a high quality. This is partly owed to the resulting rotation speeds generally having lower amplitudes.  Such significant deviations from V01's luminous TF relation seems unlikely.

\begin{figure}
\includegraphics[angle=0,width=8.5cm]{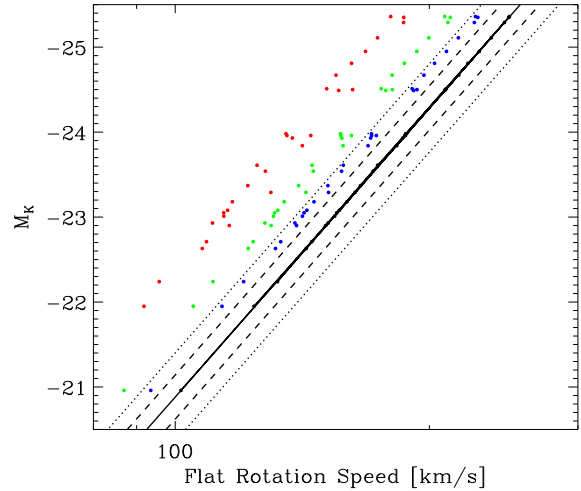}
\caption{Luminous TF relation from V01 (Eq~\ref{eqn:tfr}) showing K-band magnitude against outer rotation velocity. The solid black line is the best fit luminous TF relation and the black data points represent the DMS galaxies. The blue, green and red points are the positions those galaxies would take if their inclinations were increased by a factor 1.1, 1.2 or 1.4 respectively. The dashed and dotted lines are the 1$\sigma$ and 2$\sigma$ scatter in the observed luminous TF relation.}
\label{fig:tfrv01}
\end{figure}

To demonstrate this last point, in Fig~\ref{fig:tfrv01} the luminous TF relation from V01 (Eq~\ref{eqn:tfr}) is plotted, which is used for the DMS and this study. The solid black line is the best fit TF relation and the black data points are the values of the DMS galaxies. The blue, green and red points are the positions of those galaxies if their inclinations were increased by a factor 1.1, 1.2 or 1.4 respectively. The dashed and dotted lines are the 1$\sigma$ and 2$\sigma$ scatter in the observed TF relation. Clearly even changing the inclinations by a factor of 1.1 would be in stark conflict with the TF relation.

\subsection{Variable MOND acceleration parameter}
\begin{figure}
\includegraphics[angle=0,width=8.50cm]{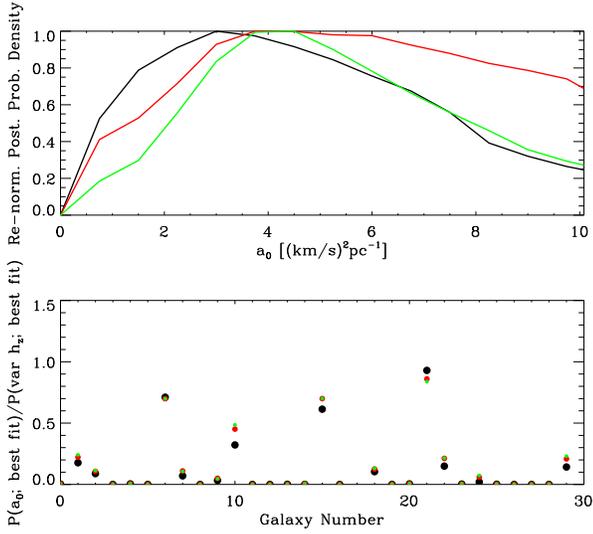}
\caption{(top panel) Co-added re-normalised posterior probability density for all 30 galaxies, this time against the MOND acceleration parameter, $a_0$. Three interpolating functions were used: $\gamma=1$, 4 and 8 (black, red and green lines respectively). The default value for the MOND acceleration parameter is $a_0=3.6~(\kms)^2\rm pc^{-1}$. (bottom panel) Ratio between the posterior probability of the best fit using a variable $a_0$ and the benchmark fit. The three interpolating functions are represented by the same colours as the lines in the top panel. Varying $a_0$ does not allow good fits to the DMS data.}
\label{fig:a0}
\end{figure}

The DMS data seem to be at odds with MOND when using the default acceleration parameter $a_0=3.6~(\kms)^2\rm pc^{-1}$. To test this parameter's influence, models were tested where the scale-height and inclination are fixed, but the MOND acceleration parameter is variable - along with the $M/L$. Although varying the MOND acceleration parameter from galaxy to galaxy is not an acceptable solution, this could identify a unique value for $a_0$ that is more consistent with the DMS data. In the top panel of Fig~\ref{fig:a0} another co-added, posterior probability density is plotted, this time against the MOND acceleration parameter. This is done for three different interpolating functions ($\gamma=1$, 4 and 8 - black, red and green lines respectively). The parameters $\gamma=4$ and 8 are trialled to see if the sharper transition from the MOND regime to the Newtonian regime would affect the ability to fit the DMS data. It appears that little is gained overall by varying $a_0$ and the same is true on a case by case basis. This can be demonstrated by comparing the fit with $a_0$ free (and scale-height fixed) to the benchmark fit (similar to what was done in \S\ref{sec:vsve} and Fig~\ref{fig:alpbet} bottom panel). In the bottom panel of Fig~\ref{fig:a0}, for each galaxy individually, the ratio is plotted between the posterior probability of the best fit using a variable $a_0$ (with fixed scale-height) and the benchmark fit. The vast majority of the galaxies would have poor fits with a variable $a_0$ and fixed scale-height.

Varying $a_0$ suffers from the same drawbacks as varying the $M/L$. Increasing $a_0$ can allow a fit to the rotation curve with a lower $M/L$, but this simultaneously increases the vertical gravitational field, which in turn requires the scale-height to be reduced.

In the second (with $\gamma=8$) and third last (with $\gamma=4$) columns of Table~\ref{tab:sims} the $M/L$ range is displayed for each galaxy individually when the MOND acceleration parameter is variable.

\subsection{$Sech^2$ vertical distributions}
\begin{figure}
\includegraphics[angle=0,width=8.50cm]{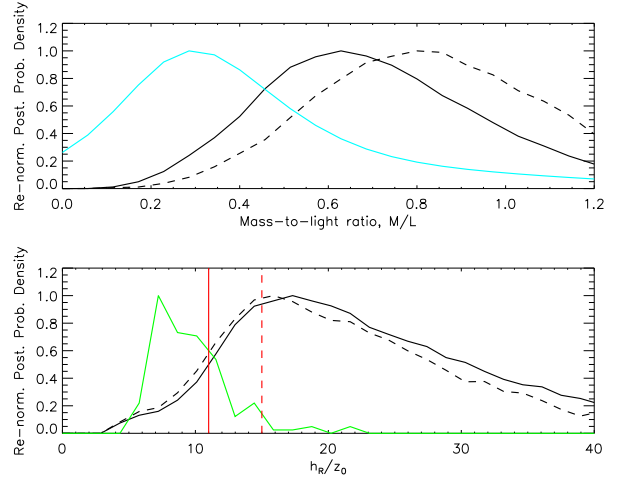}
\caption{(top panel) Superposition of the re-normalised posterior probability densities of fitted mass-to-light ratios, using a $sech^2$ function to describe the vertical distribution of the stellar disk, for the 30 galaxies in the DiskMass sample. The black solid curve corresponds to the $\gamma=1$ interpolating function and the black dashed curve to $\gamma=2$. The distribution found by the DMS, with an exponential vertical distribution, uses a turquoise curve and is just shown for reference.
(bottom panel) Superposition of the re-normalised posterior probability density of fitted scale-length to vertical scale-height, using a $sech^2$ distribution, for the 30 galaxies in the DiskMass sample. The black solid curve corresponds to the $\gamma=1$ interpolating function and the black dashed curve to $\gamma=2$. The green curve is the probability density of the sample of 153 galaxies from Bizyaev and Mitronova (2002). The vertical solid red line represents the value of $h_R/z_0$ at which the sample of Mosenkov et al. (2010) begins to decline from a previously flat value. The dashed red line represents where the probability density of Mosenkov et al. (2010) drops to zero. Using a $sech^2$ distribution does not improve the ability of MOND to fit the DMS without decreased scale-heights.}
\label{fig:sech}
\end{figure}

\begin{figure*}
\includegraphics[angle=0,width=18.0cm]{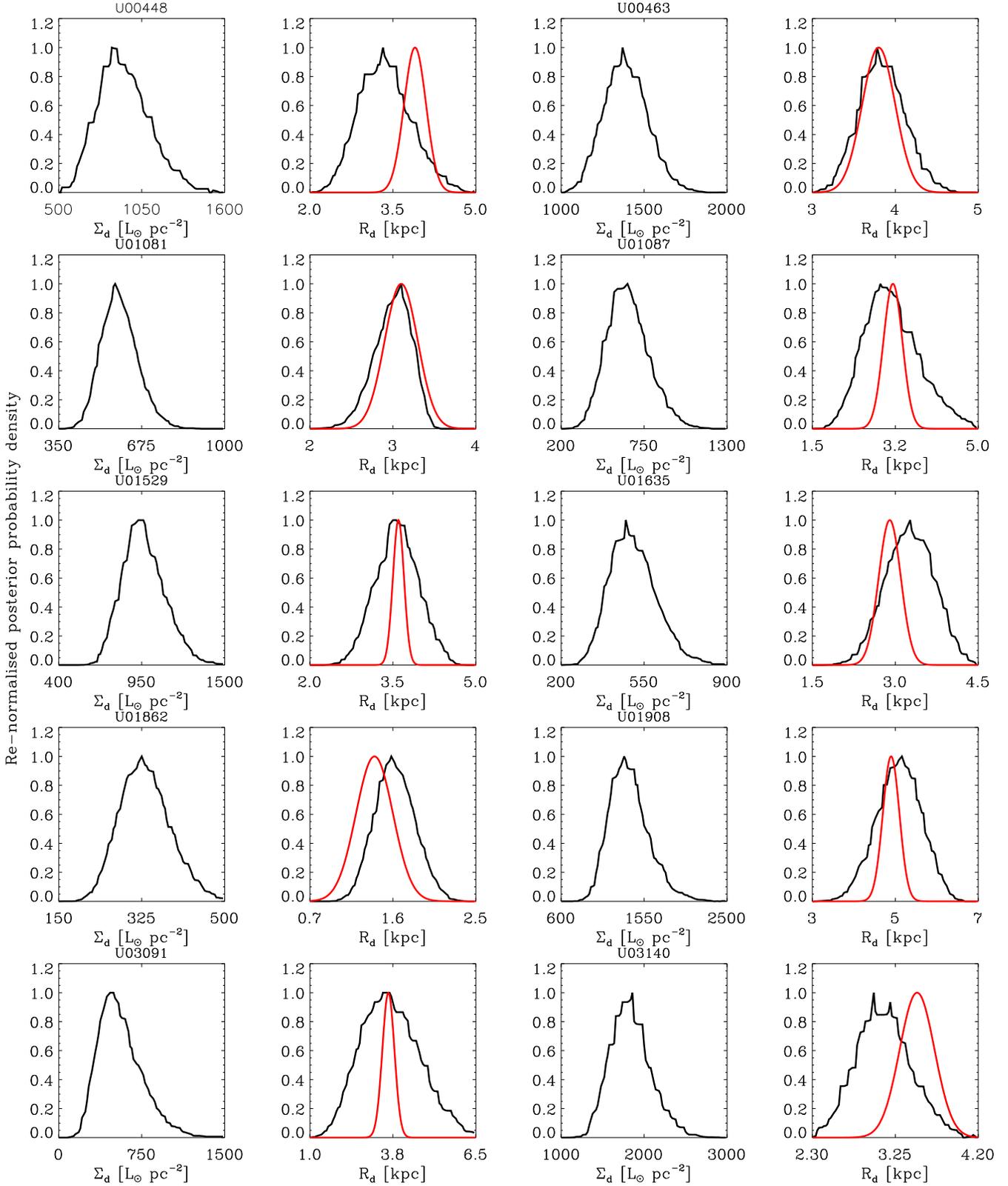}
\caption{Re-normalised posterior probability profiles for the two disk surface brightness parameters (Eq~\ref{eqn:sigdisk}) fitted to the DMS galaxy surface brightness profiles (DMSvii). In the first and third columns the disk central surface brightness is plotted and the disk scale-length is plotted in the second and fourth columns. For the disk scale-length, the values found by DMSvi (red curves) are over-plotted for comparison. }
\label{fig:like0}
\end{figure*}

\protect\label{sec:sech}
Regarding whether the poor results were caused by using an exponential function to describe the vertical scale-height, the DMS data was modelled using a $sech^2(z/z_0)$ function (\citealt{vanderkruit81a}) to describe the vertical distribution of the stellar disk in the N-particle models. The scale-height used in the $sech^2$ function is $z_0$. In Fig~\ref{fig:sech} the posterior probability densities are plotted for the ratios of the fitted $h_R$ to $z_0$ for the $\gamma=1$ and 2 interpolating functions. These are compared with data from \cite{mosenkov10} and \cite{bizyaev02}. From Fig~5d of \cite{mosenkov10} the indicative range of observed $h_R/z_0$ is between 12 and 22. A solid vertical red line is placed in Fig~\ref{fig:sech} to mark the $h_R/z_0$ value where their histogram begins to decline and a dashed vertical red line to mark where their data effectively terminates.

The 153 galaxies of \cite{bizyaev02} are also used to produce a probability density (green curve in Fig~\ref{fig:sech}) to compare with the posterior probability density found here. The $h_R/z_0$ distributions (solid and dashed black lines in Fig~\ref{fig:sech}) fitted here peak beyond where \cite{mosenkov10} and \cite{bizyaev02} have any evidence of such values. The distribution found here also extends to much larger values. To demonstrate the mismatch statistically, a 2D KS test was performed to compare the $h_R$ and $z_0$ distributions found here with the data of \cite{bizyaev02}. The confidence levels found for the $\gamma=1$ and 2 interpolating functions were $6.0\times10^{-7}$ and $2.8\times10^{-6}$. Evidently, the $sech^2$ vertical distribution does little to aid the agreement between MOND and the DMS data.

\section{Conclusions and Discussion}
By assuming the scale-heights of face-on stellar disks are similar to the measured scale-heights of edge-on galaxies (DMSii), the DMS found that stellar disks are sub-maximal, with K-band mass-to-light ratios of the order of $M/L \simeq 0.3$ (DMSvi).

In this paper the measured vertical velocity dispersions and rotation curves from DMSvi were analysed in the context of MOND. The problem is that the velocity dispersions measured by the DiskMass Survey can only allow for thick enough disks if the vertical gravitational field is weak. In Newtonian gravity this can be accommodated by lowering the $M/L$. In MOND, the $M/L$ must be larger to account for the rotation curve in the absence of DM. This means the vertical gravitational field is strong. Moreover, it is amplified by MOND relative to the vertical gravitational field of a disk with the same $M/L$ in Newtonian gravity.

Thus, the most straight-forward way to reduce the model vertical velocity dispersions in MOND is to decrease the stellar scale-height. If these were reduced from the values derived from observations of edge-on galaxies by roughly a factor of 2 then the DMS vertical velocity dispersions and rotation curves would be compatible with MOND. Regardless of stellar scale-length, these disks would have scale-heights between 200 and 400~pc. According to a two-dimensional K-S tests, such thin disks are strongly at odds with observations.

All other key parameters involved in analysing the DMS data in MOND were modelled, such as the stellar velocity dispersion ellipsoid parameters, inclination, the choice of stellar vertical distribution profile, the MOND acceleration parameter and interpolating function. No clear way was found to reconcile MOND with the data.

If the derived rotation velocities were lowered, by imposing 30-40\% higher inclinations, it became possible to simultaneously fit the low vertical velocity dispersions and low rotation speeds with low values for the stellar $M/L$. By design, since the outer rotation speeds are fitted, this is in accordance with the MONDian baryonic TF relation (with logarithmic slope fixed to 4, Eq~\ref{eqn:btfr}). Despite this, these circular velocities are inconsistent with the {\it luminous} TF relation which is quite well established for high-inclination galaxies. To this end, it would be interesting to make a fully self-consistent analysis that models the effect of the varying inclination on the two dimensional velocity fields and the surface brightness profiles.

A prediction was made for the scaling of the central dynamical surface density versus central surface luminosity density in MOND and compared with the DMS galaxies (see \citealt{swaters14}). MOND correctly predicts that the low surface luminosity density galaxies deviate from the simple relation $\Sigma_{dyn}(0)=0.3\msun/\lsun I_*(0)$.

In addition to viewing MOND as a modification of the law of gravity, \cite{milgrom11} has maintained that a modification of inertia might be better positioned to explain all available galactic dynamics data. In such a theory, the orbital trajectories of particles would have different inertias meaning the interpolating function in the radial direction would differ from the vertical directions. This is an intriguing possibility since the problem with MOND outlined here is that for the radial force defined by a galaxy, the corresponding vertical force is too large. It is interesting therefore that the vertical velocity dispersions of disk galaxies must increase to agree with MOND, but the observed radial velocity dispersions of some dwarf spheroidal galaxies (\citealt{angus08,angus14}) must decrease. This anecdotal evidence should be followed up with more rigorous study.

Something worth further investigation is the impact from super-thin disks, much thinner than the observed scale-heights represented here (\citealt{srook12,srook14}), on the measurement of vertical velocity dispersions by the DMS survey. Specifically, how prominent are super-thin disks in all disk galaxies,  and are the vertical velocity dispersions that the DMS measures meaningfully influenced by them? The super-thin disk found by \cite{srook14} was truncated at around 3~kpc, suggesting the influence is minimal. As discussed in DMSi, it is expected that stars in or near the mid-plane diffuse over time. The youngest stars may contribute to the light because of the OB stars, but they contribute little to the absorption lines, because those OB stars have weak spectral features (except for H and He).

It is possible to make even stronger claims about the status of MOND-like theories, as well as conventional dynamics, with certain desiderata. From a theoretical point of view, this would be a detailed study of the impact of the neglected cross-term (tilt of the velocity ellipsoid) in Eq~\ref{eqn:sigred}. From an observational point of view, a larger sample of edge-on galaxies (like K02) with near-infrared photometry, to more precisely confirm the correlation between scale-length and scale-height. It would also be beneficial to have an increased sample of nearly face-on galaxies, with measured stellar velocity dispersions and rotation curves and with greater sensitivity to be able to measure the stellar velocity dispersions to larger radii.

\section{acknowledgements} GWA is a postdoctoral fellow of the FWO Vlaanderen (Belgium). AD acknowledges partial support from the
INFN grant Indark, the grant Progetti di Ateneo TO Call 2012 0011 ‘Marco Polo’
of the University of Torino and the grant
PRIN 2012 ``Fisica Astroparticellare Teorica'' of the Italian Ministry of University and Research. The authors thank M. Bershady, K. Westfall and F. Lelli for valuable comments and are deeply indebted to the DiskMass Survey team, and Thomas Martinsson in particular, for taking and sharing their data. The reviewer also suggested several important improvements to the article.

%\bibliography{fgc}

\begin{thebibliography}{74}
\expandafter\ifx\csname natexlab\endcsname\relax\def\natexlab#1{#1}\fi

\bibitem[{{Andersen} \& {Bershady}(2013)}]{andersen13}
{Andersen} D.~R., {Bershady} M.~A., 2013, \apj, 768, 41

\bibitem[{{Angus}(2008)}]{angus08}
{Angus} G.~W., 2008, \mnras, 387, 1481

\bibitem[{{Angus} {et~al.}(2014){Angus}, {Gentile}, {Diaferio}, {Famaey}, \&
  {Heyden}}]{angus14}
{Angus} G.~W., {Gentile} G., {Diaferio} A., {Famaey} B., {Heyden} K.~J.~v.~d.,
  2014, \mnras, 440, 746

\bibitem[{{Angus} {et~al.}(2012){Angus}, {van der Heyden}, {Famaey}, {Gentile},
  {McGaugh}, \& {de Blok}}]{angus12}
{Angus} G.~W., {van der Heyden} K.~J., {Famaey} B., {Gentile} G., {McGaugh}
  S.~S., {de Blok} W.~J.~G., 2012, \mnras, 421, 2598

\bibitem[{{Bahcall}(1984)}]{bahcall84}
{Bahcall} J.~N., 1984, \apj, 276, 156

\bibitem[{{Begeman}(1989)}]{begeman89}
{Begeman} K.~G., 1989, \aap, 223, 47

\bibitem[{{Bell} \& {de Jong}(2001)}]{belldejong}
{Bell} E.~F., {de Jong} R.~S., 2001, \apj, 550, 212

\bibitem[{{Bell} {et~al.}(2007){Bell}, {Zheng}, {Papovich}, {Borch}, {Wolf}, \&
  {Meisenheimer}}]{bell07}
{Bell} E.~F., {Zheng} X.~Z., {Papovich} C., {Borch} A., {Wolf} C.,
  {Meisenheimer} K., 2007, \apj, 663, 834

\bibitem[{{Bershady} {et~al.}(2011){Bershady}, {Martinsson}, {Verheijen},
  {Westfall}, {Andersen}, \& {Swaters}}]{bershady11}
{Bershady} M.~A., {Martinsson} T.~P.~K., {Verheijen} M.~A.~W., {Westfall}
  K.~B., {Andersen} D.~R., {Swaters} R.~A., 2011, \apjl, 739, L47

\bibitem[{{Bershady} {et~al.}(2010{\natexlab{a}}){Bershady}, {Verheijen},
  {Swaters}, {Andersen}, {Westfall}, \& {Martinsson}}]{bershady10a}
{Bershady} M.~A., {Verheijen} M.~A.~W., {Swaters} R.~A., {Andersen} D.~R.,
  {Westfall} K.~B., {Martinsson} T., 2010{\natexlab{a}}, \apj, 716, 198 (DMSi)

\bibitem[{{Bershady} {et~al.}(2010{\natexlab{b}}){Bershady}, {Verheijen},
  {Westfall}, {Andersen}, {Swaters}, \& {Martinsson}}]{bershady10b}
{Bershady} M.~A., {Verheijen} M.~A.~W., {Westfall} K.~B., {Andersen} D.~R.,
  {Swaters} R.~A., {Martinsson} T., 2010{\natexlab{b}}, \apj, 716, 234 (DMSii)

\bibitem[{{Bienaym{\'e}} {et~al.}(2009){Bienaym{\'e}}, {Famaey}, {Wu}, {Zhao},
  \& {Aubert}}]{bienayme09}
{Bienaym{\'e}} O., {Famaey} B., {Wu} X., {Zhao} H.~S., {Aubert} D., 2009, \aap,
  500, 801

\bibitem[{{Binney} \& {Merrifield}(1998)}]{binneymerri98}
{Binney} J., {Merrifield} M., 1998, {Galactic Astronomy}

\bibitem[{{Bissantz} \& {Gerhard}(2002)}]{bissantz02}
{Bissantz} N., {Gerhard} O., 2002, \mnras, 330, 591

\bibitem[{{Bizyaev} \& {Mitronova}(2002)}]{bizyaev02}
{Bizyaev} D., {Mitronova} S., 2002, \aap, 389, 795

\bibitem[{{Blanton} {et~al.}(2003){Blanton}, {Brinkmann}, {Csabai}, {Doi},
  {Eisenstein}, {Fukugita}, {Gunn}, {Hogg}, \& {Schlegel}}]{blanton03}
{Blanton} M.~R., {Brinkmann} J., {Csabai} I., {Doi} M., {Eisenstein} D.,
  {Fukugita} M., {Gunn} J.~E., {Hogg} D.~W., {Schlegel} D.~J., 2003, \aj, 125,
  2348

\bibitem[{{Bosma}(1978)}]{bosma78}
{Bosma} A., 1978, PhD thesis, PhD Thesis, Groningen Univ., (1978)

\bibitem[{{Bosma}(1981{\natexlab{a}})}]{bosma81a}
---, 1981{\natexlab{a}}, \aj, 86, 1791

\bibitem[{{Bosma}(1981{\natexlab{b}})}]{bosma81b}
---, 1981{\natexlab{b}}, \aj, 86, 1825

\bibitem[{{Bottema}(1993)}]{bottema93}
{Bottema} R., 1993, \aap, 275, 16

\bibitem[{{Chabrier}(2003)}]{chabrier03}
{Chabrier} G., 2003, PASP, 115, 763

\bibitem[{{Conroy} \& {Gunn}(2010)}]{conroy10b}
{Conroy} C., {Gunn} J.~E., 2010, \apj, 712, 833

\bibitem[{{Conroy} {et~al.}(2009){Conroy}, {Gunn}, \& {White}}]{conroy09}
{Conroy} C., {Gunn} J.~E., {White} M., 2009, \apj, 699, 486

\bibitem[{{Conroy} {et~al.}(2010){Conroy}, {White}, \& {Gunn}}]{conroy10a}
{Conroy} C., {White} M., {Gunn} J.~E., 2010, \apj, 708, 58

\bibitem[{{Courteau} \& {Rix}(1999)}]{courteau99}
{Courteau} S., {Rix} H.-W., 1999, \apj, 513, 561

\bibitem[{{de Blok}(2010)}]{deblok10}
{de Blok} W.~J.~G., 2010, Advances in Astronomy, 2010

\bibitem[{{de Blok} \& {McGaugh}(1998)}]{deblok98}
{de Blok} W.~J.~G., {McGaugh} S.~S., 1998, \apj, 508, 132

\bibitem[{{de Blok} {et~al.}(2001){de Blok}, {McGaugh}, {Bosma}, \&
  {Rubin}}]{deblok01}
{de Blok} W.~J.~G., {McGaugh} S.~S., {Bosma} A., {Rubin} V.~C., 2001, \apjl,
  552, L23

\bibitem[{{de Blok} {et~al.}(2008){de Blok}, {Walter}, {Brinks},
  {Trachternach}, {Oh}, \& {Kennicutt}}]{deblok08}
{de Blok} W.~J.~G., {Walter} F., {Brinks} E., {Trachternach} C., {Oh} S.-H.,
  {Kennicutt} Jr. R.~C., 2008, \aj, 136, 2648

\bibitem[{{Dutton} {et~al.}(2011){Dutton}, {Brewer}, {Marshall}, {Auger},
  {Treu}, {Koo}, {Bolton}, {Holden}, \& {Koopmans}}]{dutton11}
{Dutton} A.~A., {Brewer} B.~J., {Marshall} P.~J., {Auger} M.~W., {Treu} T.,
  {Koo} D.~C., {Bolton} A.~S., {Holden} B.~P., {Koopmans} L.~V.~E., 2011,
  \mnras, 417, 1621

\bibitem[{{Famaey} \& {McGaugh}(2012)}]{famaey12}
{Famaey} B., {McGaugh} S.~S., 2012, Living Reviews in Relativity, 15, 10

\bibitem[{{Flores} \& {Primack}(1994)}]{flores94}
{Flores} R.~A., {Primack} J.~R., 1994, \apjl, 427, L1

\bibitem[{{Gentile} {et~al.}(2004){Gentile}, {Salucci}, {Klein}, {Vergani}, \&
  {Kalberla}}]{gentile04}
{Gentile} G., {Salucci} P., {Klein} U., {Vergani} D., {Kalberla} P., 2004,
  \mnras, 351, 903

\bibitem[{{Gentile} {et~al.}(2015){Gentile}, {Tydtgat}, {Baes}, {De Geyter},
  {Koleva}, {Angus}, {de Blok}, {Saftly}, \& {Viaene}}]{gentile15}
{Gentile} G., {Tydtgat} C., {Baes} M., {De Geyter} G., {Koleva} M., {Angus}
  G.~W., {de Blok} W.~J.~G., {Saftly} W., {Viaene} S., 2015, ArXiv:1502.04716

\bibitem[{{Gerssen} \& {Shapiro Griffin}(2012)}]{gerssen12}
{Gerssen} J., {Shapiro Griffin} K., 2012, \mnras, 423, 2726

\bibitem[{{Gilmore} {et~al.}(2007){Gilmore}, {Wilkinson}, {Wyse}, {Kleyna},
  {Koch}, {Evans}, \& {Grebel}}]{gilmore07b}
{Gilmore} G., {Wilkinson} M.~I., {Wyse} R.~F.~G., {Kleyna} J.~T., {Koch} A.,
  {Evans} N.~W., {Grebel} E.~K., 2007, \apj, 663, 948

\bibitem[{{Herrmann} \& {Ciardullo}(2009)}]{herrmann09}
{Herrmann} K.~A., {Ciardullo} R., 2009, \apj, 705, 1686

\bibitem[{{Kregel} {et~al.}(2002){Kregel}, {van der Kruit}, \& {de
  Grijs}}]{kregel02}
{Kregel} M., {van der Kruit} P.~C., {de Grijs} R., 2002, \mnras, 334, 646 (K02)

\bibitem[{{Kroupa}(2001)}]{kroupa01}
{Kroupa} P., 2001, \mnras, 322, 231

\bibitem[{{Kuijken} \& {Gilmore}(1989)}]{kuijken89}
{Kuijken} K., {Gilmore} G., 1989, \mnras, 239, 571

\bibitem[{{Kuijken} \& {Gilmore}(1991)}]{kuijken91}
---, 1991, \apjl, 367, L9

\bibitem[{{Martinsson} {et~al.}(2013{\natexlab{a}}){Martinsson}, {Verheijen},
  {Westfall}, {Bershady}, {Andersen}, \& {Swaters}}]{martinsson13b}
{Martinsson} T.~P.~K., {Verheijen} M.~A.~W., {Westfall} K.~B., {Bershady}
  M.~A., {Andersen} D.~R., {Swaters} R.~A., 2013{\natexlab{a}}, \aap, 557, A131
  (dmsvii)

\bibitem[{{Martinsson} {et~al.}(2013{\natexlab{b}}){Martinsson}, {Verheijen},
  {Westfall}, {Bershady}, {Schechtman-Rook}, {Andersen}, \&
  {Swaters}}]{martinsson13a}
{Martinsson} T.~P.~K., {Verheijen} M.~A.~W., {Westfall} K.~B., {Bershady}
  M.~A., {Schechtman-Rook} A., {Andersen} D.~R., {Swaters} R.~A.,
  2013{\natexlab{b}}, \aap, 557, A130 (DMSvi)

\bibitem[{{McGaugh} \& {Schombert}(2015)}]{mcgaugh15}
{McGaugh} S., {Schombert} J., 2015, ArXiv:1501.06826

\bibitem[{{McGaugh}(2005)}]{mcgaugh05a}
{McGaugh} S.~S., 2005, \apj, 632, 859

\bibitem[{{McGaugh} \& {Schombert}(2014)}]{mcgaugh14}
{McGaugh} S.~S., {Schombert} J.~M., 2014, \aj, 148, 77

\bibitem[{{McGaugh} {et~al.}(2000){McGaugh}, {Schombert}, {Bothun}, \& {de
  Blok}}]{mcgaugh00}
{McGaugh} S.~S., {Schombert} J.~M., {Bothun} G.~D., {de Blok} W.~J.~G., 2000,
  \apj, 533, L99

\bibitem[{{Milgrom}(1983)}]{milgrom83a}
{Milgrom} M., 1983, \apj, 270, 365

\bibitem[{{Milgrom}(2010)}]{milgrom10}
---, 2010, \mnras, 403, 886

\bibitem[{{Milgrom}(2011)}]{milgrom11}
---, 2011, ArXiv:1111.1611

\bibitem[{{Mosenkov} {et~al.}(2010){Mosenkov}, {Sotnikova}, \&
  {Reshetnikov}}]{mosenkov10}
{Mosenkov} A.~V., {Sotnikova} N.~Y., {Reshetnikov} V.~P., 2010, \mnras, 401,
  559

\bibitem[{{Nipoti} {et~al.}(2007){Nipoti}, {Londrillo}, {Zhao}, \&
  {Ciotti}}]{nipoti07b}
{Nipoti} C., {Londrillo} P., {Zhao} H.~S., {Ciotti} L., 2007, \mnras, 379, 597

\bibitem[{{Pohlen} {et~al.}(2000){Pohlen}, {Dettmar}, {L{\"u}tticke}, \&
  {Schwarzkopf}}]{pohlen00}
{Pohlen} M., {Dettmar} R.-J., {L{\"u}tticke} R., {Schwarzkopf} U., 2000, \aap,
  144, 405

\bibitem[{{Press} {et~al.}(1992){Press}, {Teukolsky}, {Vetterling}, \&
  {Flannery}}]{press92}
{Press} W.~H., {Teukolsky} S.~A., {Vetterling} W.~T., {Flannery} B.~P., 1992,
  {Numerical recipes in FORTRAN. The art of scientific computing}

\bibitem[{{Rubin} {et~al.}(1978){Rubin}, {Thonnard}, \& {Ford}}]{rubin78}
{Rubin} V.~C., {Thonnard} N., {Ford} J. W.~K., 1978, \apjl, 225, L107

\bibitem[{{Sackett}(1997)}]{sackett97}
{Sackett} P.~D., 1997, \apj, 483, 103

\bibitem[{{Schechtman-Rook} \& {Bershady}(2014)}]{srook14}
{Schechtman-Rook} A., {Bershady} M.~A., 2014, ArXiv:1409.6727

\bibitem[{{Schechtman-Rook} {et~al.}(2012){Schechtman-Rook}, {Bershady}, \&
  {Wood}}]{srook12}
{Schechtman-Rook} A., {Bershady} M.~A., {Wood} K., 2012, \apj, 746, 70

\bibitem[{{Schwarzkopf} \& {Dettmar}(2000)}]{schwarzkopf00}
{Schwarzkopf} U., {Dettmar} R.-J., 2000, \aap, 144, 85

\bibitem[{{Sellwood} \& {Debattista}(2014)}]{sellwood14}
{Sellwood} J.~A., {Debattista} V.~P., 2014, ArXiv:1410.0834

\bibitem[{{Sofue} \& {Rubin}(2001)}]{sofue01}
{Sofue} Y., {Rubin} V., 2001, \araa, 39, 137

\bibitem[{{Swaters} {et~al.}(2014){Swaters}, {Bershady}, {Martinsson},
  {Westfall}, {Andersen}, \& {Verheijen}}]{swaters14}
{Swaters} R.~A., {Bershady} M.~A., {Martinsson} T.~P.~K., {Westfall} K.~B.,
  {Andersen} D.~R., {Verheijen} M.~A.~W., 2014, \apjl, 797, L28

\bibitem[{{Swaters} {et~al.}(2003){Swaters}, {Madore}, {van den Bosch}, \&
  {Balcells}}]{swaters03}
{Swaters} R.~A., {Madore} B.~F., {van den Bosch} F.~C., {Balcells} M., 2003,
  \apj, 583, 732

\bibitem[{{van Albada} {et~al.}(1985){van Albada}, {Bahcall}, {Begeman}, \&
  {Sancisi}}]{vanalbada85}
{van Albada} T.~S., {Bahcall} J.~N., {Begeman} K., {Sancisi} R., 1985, \apj,
  295, 305

\bibitem[{{van Albada} \& {Sancisi}(1986)}]{vanalbada86}
{van Albada} T.~S., {Sancisi} R., 1986, Royal Society of London Philosophical
  Transactions Series A, 320, 447

\bibitem[{{van den Bosch} \& {Swaters}(2001)}]{vandenbosch01}
{van den Bosch} F.~C., {Swaters} R.~A., 2001, \mnras, 325, 1017

\bibitem[{{van der Hulst} {et~al.}(1992){van der Hulst}, {Terlouw}, {Begeman},
  {Zwitser}, \& {Roelfsema}}]{vdhulst92}
{van der Hulst} J.~M., {Terlouw} J.~P., {Begeman} K.~G., {Zwitser} W.,
  {Roelfsema} P.~R., 1992, in Astronomical Society of the Pacific Conference
  Series, Vol.~25, Astronomical Data Analysis Software and Systems I, {Worrall}
  D.~M., {Biemesderfer} C., {Barnes} J., eds., p. 131

\bibitem[{{van der Kruit} \& {Searle}(1981)}]{vanderkruit81a}
{van der Kruit} P.~C., {Searle} L., 1981, \aap, 95, 105

\bibitem[{{van der Kruit} \& {Searle}(1982)}]{vanderkruit82}
---, 1982, \aap, 110, 61

\bibitem[{{Verheijen}(2001)}]{verheijen01}
{Verheijen} M.~A.~W., 2001, \apj, 563, 694

\bibitem[{{Westfall}(2015)}]{westfall15}
{Westfall} K.~B., 2015, in prep

\bibitem[{{Westfall} {et~al.}(2011){Westfall}, {Bershady}, {Verheijen},
  {Andersen}, {Martinsson}, {Swaters}, \& {Schechtman-Rook}}]{westfall11}
{Westfall} K.~B., {Bershady} M.~A., {Verheijen} M.~A.~W., {Andersen} D.~R.,
  {Martinsson} T.~P.~K., {Swaters} R.~A., {Schechtman-Rook} A., 2011, \apj,
  742, 18 (DMSiv)

\bibitem[{{Xilouris} {et~al.}(1999){Xilouris}, {Byun}, {Kylafis}, {Paleologou},
  \& {Papamastorakis}}]{xilouris99}
{Xilouris} E.~M., {Byun} Y.~I., {Kylafis} N.~D., {Paleologou} E.~V.,
  {Papamastorakis} J., 1999, \aap, 344, 868

\bibitem[{{Xilouris} {et~al.}(1997){Xilouris}, {Kylafis}, {Papamastorakis},
  {Paleologou}, \& {Haerendel}}]{xilouris97}
{Xilouris} E.~M., {Kylafis} N.~D., {Papamastorakis} J., {Paleologou} E.~V.,
  {Haerendel} G., 1997, \aap, 325, 135

\end{thebibliography}

\begin{landscape}
\begin{table}
\centering
\begin{tabular}{|l||cc||cc|cc|cc|c|c|c|}
\hline
&\multicolumn{2}{c}{DMS} \vline \vline & \multicolumn{9}{c}{MOND} \vline \\
\hline
&\multicolumn{2}{c}{} \vline \vline& \multicolumn{6}{c}{$h_z$ and $\Upsilon_d$ free; $i$ constrained} \vline&  \multicolumn{2}{c}{$a_0$ and $\Upsilon_d$ free} \vline & $i$ and $\Upsilon_d$ free \\
\hline
&&&\multicolumn{2}{c}{$\gamma=1$}\vline&\multicolumn{2}{c}{$\gamma=2$}\vline&\multicolumn{2}{c}{\tiny $\delta=4$} \vline& $\gamma=4$ & $\gamma=8$ & $\gamma=2$\\
Galaxy& $h_{z,DMS}$ (kpc)& $\Upsilon_{d,DMS}$ & ${h_z \over h_{z,DMS}}$ &  $\Upsilon_d$ & ${h_z \over h_{z,DMS}}$ &  $\Upsilon_d$& ${h_z \over h_{z,DMS}}$ &  $\Upsilon_d$  &  $\Upsilon_d$ &  $\Upsilon_d$ &  $\Upsilon_d$ \\
\hline
U00448 &0.46$\pm$0.10&0.31$\pm$0.21&$0.50^{+0.12}_{-0.12}$&$0.59^{+0.19}_{-0.13}$&$0.50^{+0.15}_{-0.10}$&$0.71^{+0.23}_{-0.15}$&$0.53^{+0.17}_{-0.08}$&$0.48^{+0.24}_{-0.09}$&$0.55^{+0.20}_{-0.20}$&$0.60^{+0.17}_{-0.21}$&$0.11^{+0.16}_{-0.08}$\\
U00463 &0.45$\pm$0.10&0.29$\pm$0.17&$0.65^{+0.18}_{-0.12}$&$0.51^{+0.17}_{-0.12}$&$0.67^{+0.15}_{-0.13}$&$0.61^{+0.15}_{-0.12}$&$0.63^{+0.15}_{-0.12}$&$0.71^{+0.19}_{-0.15}$&$0.63^{+0.11}_{-0.11}$&$0.64^{+0.09}_{-0.08}$&$0.21^{+0.20}_{-0.17}$\\
U01081 &0.40$\pm$0.09&0.40$\pm$0.16&$0.63^{+0.17}_{-0.12}$&$0.53^{+0.17}_{-0.13}$&$0.65^{+0.18}_{-0.10}$&$0.69^{+0.21}_{-0.19}$&$0.65^{+0.15}_{-0.08}$&$0.52^{+0.19}_{-0.16}$&$0.71^{+0.17}_{-0.29}$&$0.75^{+0.16}_{-0.28}$&$0.19^{+0.19}_{-0.17}$\\
U01087 &0.41$\pm$0.09&0.32$\pm$0.13&$0.42^{+0.15}_{-0.08}$&$0.72^{+0.21}_{-0.19}$&$0.47^{+0.13}_{-0.10}$&$0.89^{+0.20}_{-0.23}$&$0.53^{+0.12}_{-0.12}$&$0.65^{+0.31}_{-0.23}$&$0.51^{+0.19}_{-0.17}$&$0.57^{+0.16}_{-0.19}$&$0.15^{+0.11}_{-0.09}$\\
U01529 &0.44$\pm$0.10&0.29$\pm$0.16&$0.40^{+0.15}_{-0.08}$&$0.67^{+0.16}_{-0.12}$&$0.43^{+0.12}_{-0.10}$&$0.87^{+0.16}_{-0.16}$&$0.40^{+0.17}_{-0.08}$&$0.89^{+0.23}_{-0.23}$&$0.91^{+0.16}_{-0.15}$&$0.92^{+0.15}_{-0.12}$&$0.21^{+0.13}_{-0.05}$\\
U01635 &0.39$\pm$0.09&0.26$\pm$0.12&$0.35^{+0.12}_{-0.08}$&$0.61^{+0.23}_{-0.13}$&$0.40^{+0.08}_{-0.10}$&$0.79^{+0.25}_{-0.17}$&$0.37^{+0.12}_{-0.05}$&$0.63^{+0.21}_{-0.19}$&$0.37^{+0.15}_{-0.16}$&$0.43^{+0.19}_{-0.19}$&$0.08^{+0.08}_{-0.07}$\\
U01862 &0.24$\pm$0.06&0.56$\pm$0.29&$0.80^{+0.30}_{-0.20}$&$0.65^{+0.32}_{-0.16}$&$0.92^{+0.30}_{-0.23}$&$0.85^{+0.27}_{-0.25}$&$0.73^{+0.12}_{-0.10}$&$0.84^{+0.15}_{-0.11}$&$0.59^{+0.01}_{-0.01}$&$0.61^{+0.03}_{-0.05}$&$0.59^{+0.37}_{-0.28}$\\
U01908 &0.53$\pm$0.12&0.26$\pm$0.18&$0.73^{+0.10}_{-0.15}$&$0.35^{+0.13}_{-0.07}$&$0.73^{+0.18}_{-0.13}$&$0.48^{+0.12}_{-0.11}$&$0.80^{+0.15}_{-0.15}$&$0.37^{+0.15}_{-0.12}$&$0.40^{+0.15}_{-0.16}$&$0.43^{+0.15}_{-0.19}$&$0.25^{+0.11}_{-0.09}$\\
U03091 &0.44$\pm$0.10&0.23$\pm$0.10&$0.35^{+0.10}_{-0.07}$&$0.80^{+0.17}_{-0.17}$&$0.38^{+0.08}_{-0.08}$&$1.01^{+0.17}_{-0.25}$&$0.38^{+0.08}_{-0.07}$&$0.85^{+0.28}_{-0.20}$&$0.48^{+0.13}_{-0.12}$&$0.52^{+0.12}_{-0.12}$&$0.09^{+0.07}_{-0.04}$\\
U03140 &0.43$\pm$0.10&0.33$\pm$0.14&$0.68^{+0.12}_{-0.12}$&$0.51^{+0.13}_{-0.11}$&$0.70^{+0.13}_{-0.13}$&$0.64^{+0.15}_{-0.13}$&$0.73^{+0.17}_{-0.10}$&$0.55^{+0.19}_{-0.07}$&$0.49^{+0.07}_{-0.05}$&$0.49^{+0.07}_{-0.04}$&$0.40^{+0.05}_{-0.04}$\\
U03701 &0.44$\pm$0.11&0.85$\pm$0.40&$0.70^{+0.23}_{-0.15}$&$0.65^{+0.27}_{-0.19}$&$0.78^{+0.22}_{-0.13}$&$0.77^{+0.28}_{-0.24}$&$0.65^{+0.25}_{-0.12}$&$0.60^{+0.32}_{-0.16}$&$0.99^{+0.65}_{-0.51}$&$1.12^{+0.60}_{-0.57}$&$0.36^{+0.37}_{-0.23}$\\
U03997 &0.58$\pm$0.14&0.45$\pm$0.19&$0.37^{+0.13}_{-0.07}$&$0.73^{+0.33}_{-0.15}$&$0.45^{+0.10}_{-0.12}$&$0.92^{+0.28}_{-0.21}$&$0.38^{+0.12}_{-0.08}$&$0.79^{+0.19}_{-0.25}$&$0.92^{+0.37}_{-0.35}$&$0.99^{+0.36}_{-0.35}$&$0.09^{+0.12}_{-0.09}$\\
U04036 &0.49$\pm$0.12&0.29$\pm$0.11&$0.45^{+0.08}_{-0.08}$&$0.60^{+0.21}_{-0.11}$&$0.45^{+0.08}_{-0.07}$&$0.87^{+0.15}_{-0.17}$&$0.43^{+0.08}_{-0.05}$&$0.93^{+0.17}_{-0.24}$&$0.48^{+0.08}_{-0.08}$&$0.52^{+0.07}_{-0.07}$&$0.21^{+0.07}_{-0.04}$\\
U04107 &0.41$\pm$0.09&0.28$\pm$0.14&$0.53^{+0.12}_{-0.12}$&$0.55^{+0.23}_{-0.12}$&$0.52^{+0.13}_{-0.08}$&$0.81^{+0.21}_{-0.21}$&$0.58^{+0.15}_{-0.12}$&$0.57^{+0.33}_{-0.23}$&$0.59^{+0.17}_{-0.43}$&$0.65^{+0.13}_{-0.21}$&$0.09^{+0.15}_{-0.09}$\\
U04256 &0.52$\pm$0.12&0.15$\pm$0.19&$0.63^{+0.08}_{-0.12}$&$0.28^{+0.12}_{-0.08}$&$0.65^{+0.12}_{-0.10}$&$0.39^{+0.16}_{-0.09}$&$0.65^{+0.10}_{-0.10}$&$0.44^{+0.13}_{-0.13}$&$0.28^{+0.09}_{-0.09}$&$0.31^{+0.08}_{-0.08}$&$0.09^{+0.08}_{-0.07}$\\
U04368 &0.41$\pm$0.10&0.57$\pm$0.34&$0.72^{+0.25}_{-0.18}$&$0.67^{+0.17}_{-0.15}$&$0.82^{+0.23}_{-0.23}$&$0.85^{+0.09}_{-0.19}$&$0.75^{+0.27}_{-0.20}$&$0.59^{+0.19}_{-0.15}$&$1.17^{+0.31}_{-0.35}$&$1.24^{+0.27}_{-0.32}$&$0.57^{+0.56}_{-0.31}$\\
U04380 &0.54$\pm$0.12&0.27$\pm$0.09&$0.45^{+0.10}_{-0.12}$&$0.55^{+0.17}_{-0.12}$&$0.48^{+0.12}_{-0.10}$&$0.65^{+0.20}_{-0.12}$&$0.48^{+0.08}_{-0.08}$&$0.59^{+0.19}_{-0.12}$&$0.35^{+0.12}_{-0.12}$&$0.41^{+0.12}_{-0.13}$&$0.25^{+0.05}_{-0.05}$\\
U04458 &0.79$\pm$0.17&0.42$\pm$0.24&$0.47^{+0.47}_{-0.18}$&$0.32^{+0.15}_{-0.08}$&$0.63^{+0.42}_{-0.30}$&$0.39^{+0.21}_{-0.07}$&$0.58^{+0.47}_{-0.27}$&$0.25^{+0.12}_{-0.07}$&$0.92^{+0.15}_{-0.13}$&$0.92^{+0.15}_{-0.12}$&$0.07^{+0.28}_{-0.05}$\\
U04555 &0.48$\pm$0.11&0.35$\pm$0.20&$0.53^{+0.17}_{-0.12}$&$0.60^{+0.13}_{-0.13}$&$0.57^{+0.18}_{-0.13}$&$0.80^{+0.17}_{-0.16}$&$0.62^{+0.20}_{-0.15}$&$0.49^{+0.21}_{-0.13}$&$1.00^{+0.19}_{-0.17}$&$1.03^{+0.17}_{-0.17}$&$0.16^{+0.20}_{-0.05}$\\
U04622 &0.70$\pm$0.16&0.28$\pm$0.17&$0.38^{+0.15}_{-0.05}$&$0.53^{+0.20}_{-0.12}$&$0.43^{+0.13}_{-0.12}$&$0.73^{+0.20}_{-0.19}$&$0.43^{+0.15}_{-0.10}$&$0.45^{+0.17}_{-0.09}$&$0.53^{+0.20}_{-0.16}$&$0.59^{+0.20}_{-0.17}$&$0.15^{+0.12}_{-0.08}$\\
U06903 &0.49$\pm$0.11&0.24$\pm$0.17&$0.27^{+0.20}_{-0.12}$&$0.92^{+0.19}_{-0.27}$&$0.33^{+0.18}_{-0.17}$&$1.04^{+0.27}_{-0.25}$&$0.32^{+0.17}_{-0.13}$&$0.83^{+0.28}_{-0.19}$&$0.91^{+0.28}_{-0.01}$&$1.16^{+0.48}_{-0.01}$&$0.13^{+0.19}_{-0.05}$\\
U06918 &0.21$\pm$0.05&0.06$\pm$0.24&$0.80^{+0.23}_{-0.20}$&$0.40^{+0.11}_{-0.13}$&$0.77^{+0.32}_{-0.17}$&$0.48^{+0.17}_{-0.09}$&$0.88^{+0.25}_{-0.23}$&$0.49^{+0.15}_{-0.16}$&$0.41^{+0.16}_{-0.20}$&$0.44^{+0.13}_{-0.17}$&$0.31^{+0.16}_{-0.12}$\\
U07244 &0.46$\pm$0.11&0.41$\pm$0.18&$0.68^{+0.18}_{-0.13}$&$0.41^{+0.20}_{-0.16}$&$0.73^{+0.18}_{-0.15}$&$0.49^{+0.21}_{-0.19}$&$0.62^{+0.18}_{-0.08}$&$0.45^{+0.23}_{-0.19}$&$0.17^{+0.27}_{-0.11}$&$0.16^{+0.35}_{-0.11}$&$0.21^{+0.08}_{-0.07}$\\
U07917 &0.76$\pm$0.17&0.29$\pm$0.33&$0.45^{+0.17}_{-0.07}$&$0.51^{+0.16}_{-0.12}$&$0.48^{+0.13}_{-0.10}$&$0.75^{+0.12}_{-0.19}$&$0.55^{+0.12}_{-0.13}$&$0.39^{+0.15}_{-0.08}$&$0.84^{+0.12}_{-0.13}$&$0.85^{+0.12}_{-0.09}$&$0.09^{+0.13}_{-0.05}$\\
U08196 &0.53$\pm$0.12&0.94$\pm$0.41&$2.18^{+0.20}_{-0.42}$&$0.49^{+0.11}_{-0.11}$&$2.10^{+0.37}_{-0.37}$&$0.68^{+0.15}_{-0.11}$&$1.55^{+0.40}_{-0.28}$&$0.95^{+0.20}_{-0.15}$&$1.00^{+0.11}_{-0.08}$&$1.01^{+0.09}_{-0.08}$&$1.39^{+0.23}_{-0.15}$\\
U09177 &0.67$\pm$0.15&0.36$\pm$0.26&$0.38^{+0.15}_{-0.10}$&$0.67^{+0.23}_{-0.11}$&$0.40^{+0.15}_{-0.08}$&$0.93^{+0.19}_{-0.20}$&$0.43^{+0.12}_{-0.12}$&$0.59^{+0.15}_{-0.13}$&$1.15^{+0.24}_{-0.20}$&$1.17^{+0.21}_{-0.19}$&$0.19^{+0.13}_{-0.05}$\\
U09837 &0.60$\pm$0.14&0.32$\pm$0.14&$0.30^{+0.10}_{-0.08}$&$0.93^{+0.36}_{-0.17}$&$0.35^{+0.10}_{-0.10}$&$1.04^{+0.39}_{-0.12}$&$0.30^{+0.12}_{-0.07}$&$0.89^{+0.31}_{-0.20}$&$0.84^{+0.39}_{-0.31}$&$1.03^{+0.28}_{-0.41}$&$0.05^{+0.12}_{-0.05}$\\
U09965 &0.44$\pm$0.10&0.25$\pm$0.12&$0.50^{+0.12}_{-0.10}$&$0.51^{+0.19}_{-0.11}$&$0.53^{+0.08}_{-0.08}$&$0.71^{+0.16}_{-0.16}$&$0.60^{+0.15}_{-0.18}$&$0.48^{+0.52}_{-0.19}$&$0.36^{+0.11}_{-0.12}$&$0.43^{+0.09}_{-0.13}$&$0.21^{+0.05}_{-0.04}$\\
U11318 &0.51$\pm$0.11&0.29$\pm$0.20&$0.73^{+0.12}_{-0.12}$&$0.39^{+0.08}_{-0.11}$&$0.78^{+0.12}_{-0.03}$&$0.49^{+0.05}_{-0.09}$&$0.70^{+0.10}_{-0.07}$&$0.57^{+0.09}_{-0.11}$&$0.37^{+0.07}_{-0.08}$&$0.41^{+0.05}_{-0.07}$&$0.32^{+0.04}_{-0.03}$\\
U12391 &0.46$\pm$0.10&0.41$\pm$0.19&$0.63^{+0.20}_{-0.13}$&$0.49^{+0.19}_{-0.11}$&$0.68^{+0.15}_{-0.12}$&$0.68^{+0.21}_{-0.15}$&$0.67^{+0.23}_{-0.13}$&$0.41^{+0.28}_{-0.09}$&$0.75^{+0.21}_{-0.28}$&$0.81^{+0.13}_{-0.23}$&$0.19^{+0.27}_{-0.17}$\\
\hline\hline
\end{tabular}
\caption{This table presents various scale-heights and mass-to-light ratios with 1$\sigma$ errors for each of the 30 galaxies in the DMS sample. Column (1) gives the galaxy name, (2) the scale-height used by the DMS, (3) has the $M/L$ fitted by the DMS using Newtonian gravity and a DM halo. The remaining columns all pertain to MOND fits. Columns (4)\&(5), (6)\&(7) and (8)\&(9) give the best fit ratio of  fitted scale-height to the one used by the DMS and the fitted $M/L$ for three distinct interpolating functions respectively. These models come from analyses where the $M/L$ and scale-height are almost unconstrained, and the inclination has a limited amount of freedom. The next two columns (10) and (11) are for models where only the MOND acceleration parameter and $M/L$ were free to vary (inclination and scale-height were fixed). These two columns use different interpolating functions and show the $M/L$ of the best fit. The last column is the best fit $M/L$ found when the inclination is unconstrained, but scale-height is fixed.}
\protect\label{tab:sims}
\end{table}
\end{landscape}

\begin{figure*}
\includegraphics[angle=0,width=18.0cm]{sbfits0.ps_pages1}
\caption{As per Fig~\ref{fig:like0}.}
\label{fig:like1}
\end{figure*}

\begin{figure*}
\includegraphics[angle=0,width=18.0cm]{sbfits0.ps_pages2}
\caption{As per Fig~\ref{fig:like0}.}
\label{fig:like2}
\end{figure*}

\clearpage

\begin{figure*}
\includegraphics[angle=0,width=18.0cm]{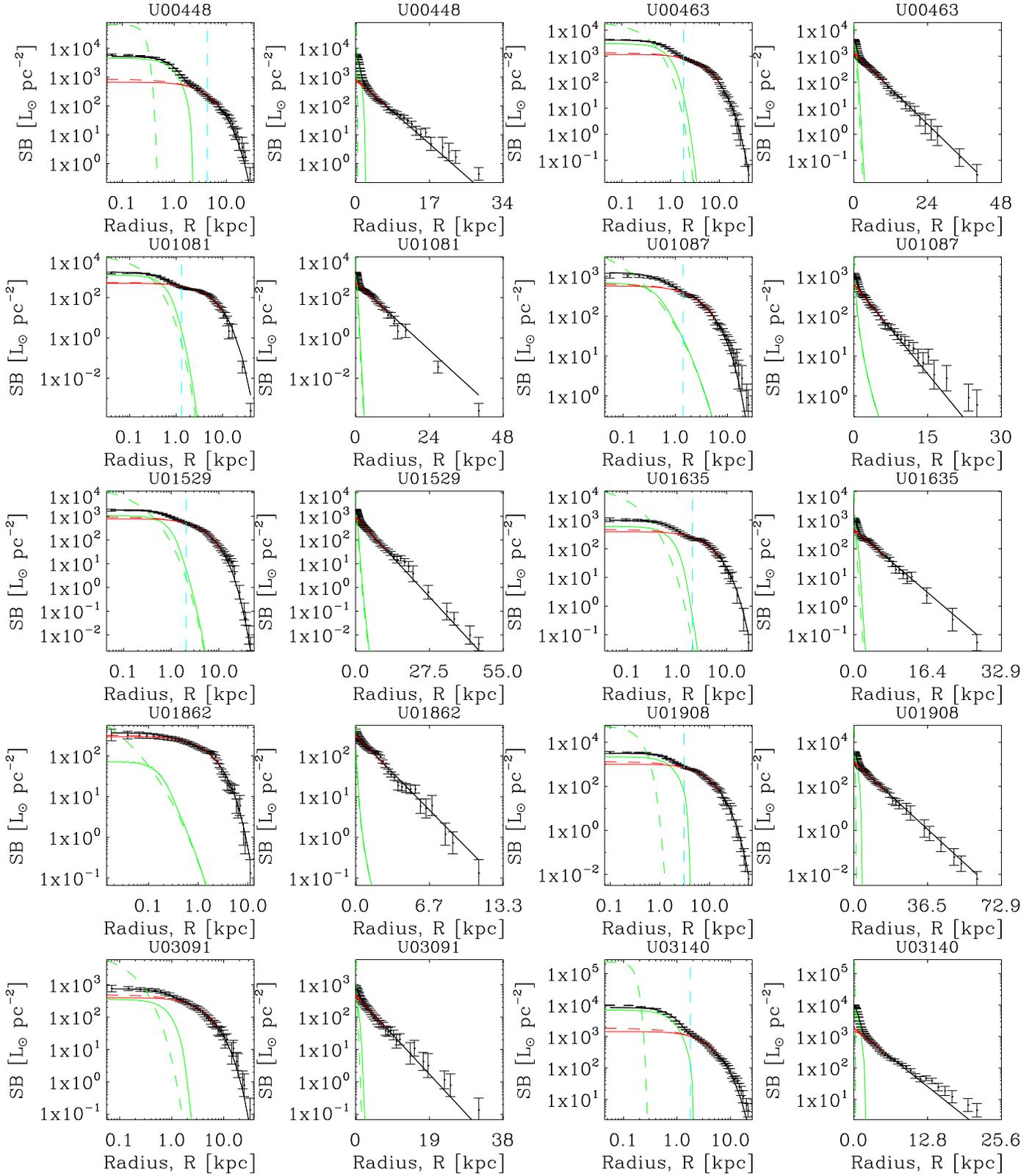}
\caption{Observed surface brightness profiles (black error bars around small circle) along the disk major axis for each galaxy in the DMS. Each row shows two different galaxies. In columns 1 and 3 the fits in log-log are plotted and in columns 2 and 4 they are plotted as log-linear, to expose the quality of the fits at different scales. The green dashed lines are the intrinsic bulge surface brightnesses and the solid green lines are the seeing affected versions. The same is true for the red dashed and solid lines, except these are for the exponential disks. The black solid lines are the seeing affected total combined surface brightnesses. The vertical turquoise lines are the bulge radii as given by DMSvi and are unchanged in this analysis.}
\label{fig:1like0}
\end{figure*}

\begin{figure*}
\includegraphics[angle=0,width=18.0cm]{sbfits1.ps_pages1}
\caption{As per Fig~\ref{fig:1like1}.}
\label{fig:1like1}
\end{figure*}

\begin{figure*}
\includegraphics[angle=0,width=18.0cm]{sbfits1.ps_pages2}
\caption{As per Fig~\ref{fig:1like1}.}
\label{fig:1like2}
\end{figure*}

\begin{figure*}
\includegraphics[angle=0,width=18.0cm]{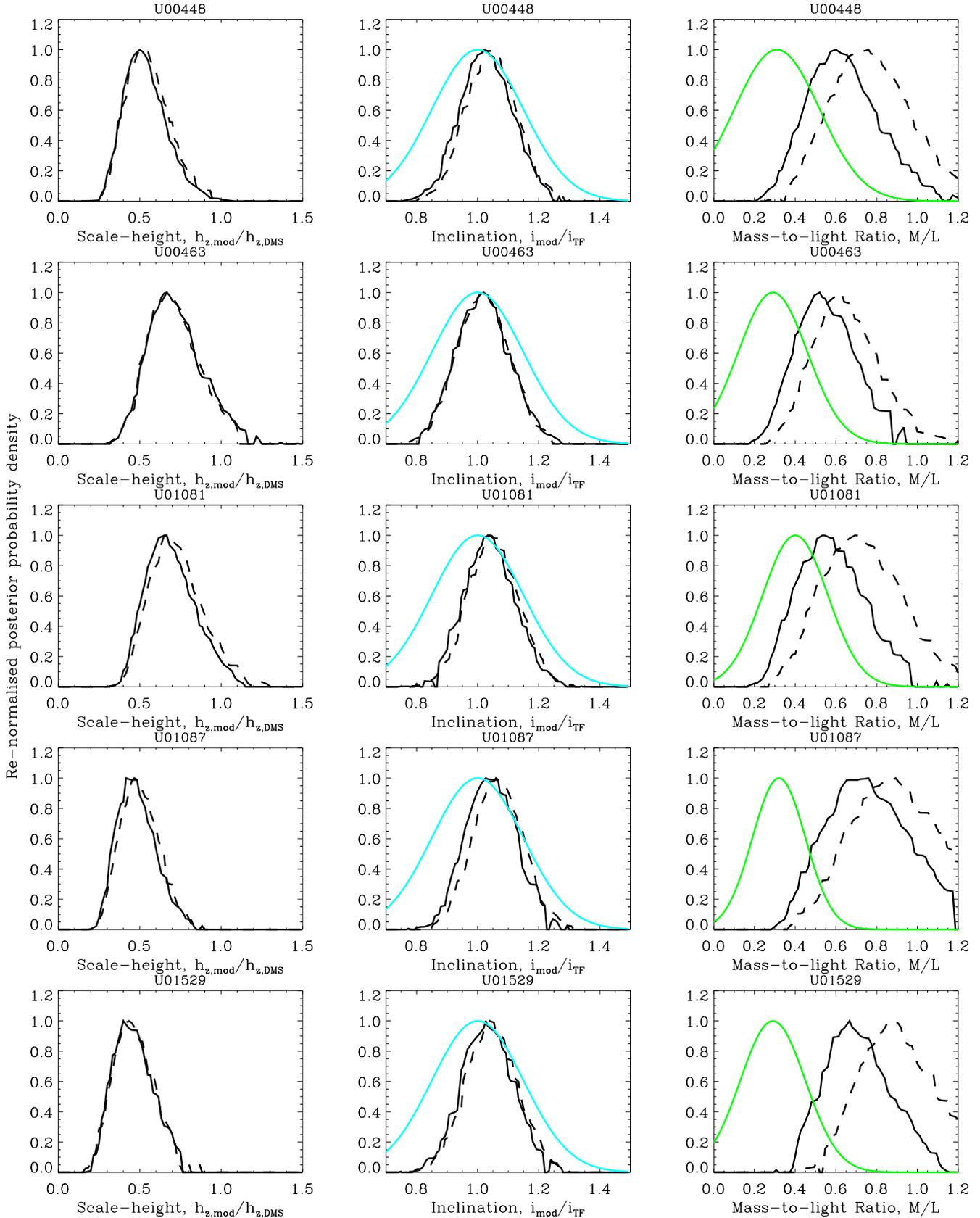}
\caption{\tiny Re-normalised posterior probability for each parameter from the fits to the vertical velocity dispersions and rotation curves. Each row displays a different galaxy. The first column shows the ratio of the fitted scale-heights to the scale-heights derived from observations of edge-on galaxies, as used by DMSvi (Eq~\ref{eqn:kreg}). The different lines are for MOND with $\gamma=1$ and 2 (solid and dashed black lines respectively). Note all rows use the same $x$-axis range except the galaxy UGC~8196. The second column shows the fitted inclination relative to the inclination from the luminous TF relation (V01). The prior on inclination is the turquoise curve. The third column shows the fitted $M/L$ and the green curve is the confidence range of the deduced $M/L$ from DMSvi (using Newtonian gravity with fitted DM halos).}
\label{fig:2like0}
\end{figure*}

\begin{figure*}
\includegraphics[angle=0,width=18.0cm]{ajamesfits.ps_pages1}
\caption{As per Fig~\ref{fig:2like0}.}
\label{fig:2like1}
\end{figure*}

\begin{figure*}
\includegraphics[angle=0,width=18.0cm]{ajamesfits.ps_pages2}
\caption{As per Fig~\ref{fig:2like0}.}
\label{fig:2like2}
\end{figure*}

\begin{figure*}
\includegraphics[angle=0,width=18.0cm]{ajamesfits.ps_pages3}
\caption{As per Fig~\ref{fig:2like0}.}
\label{fig:2like3}
\end{figure*}

\begin{figure*}
\includegraphics[angle=0,width=18.0cm]{ajamesfits.ps_pages4}
\caption{As per Fig~\ref{fig:2like0}.}
\label{fig:2like4}
\end{figure*}

\begin{figure*}
\includegraphics[angle=0,width=18.0cm]{ajamesfits.ps_pages5}
\caption{As per Fig~\ref{fig:2like0}.}
\label{fig:2like5}
\end{figure*}

\begin{figure*}
\includegraphics[angle=0,width=18.0cm]{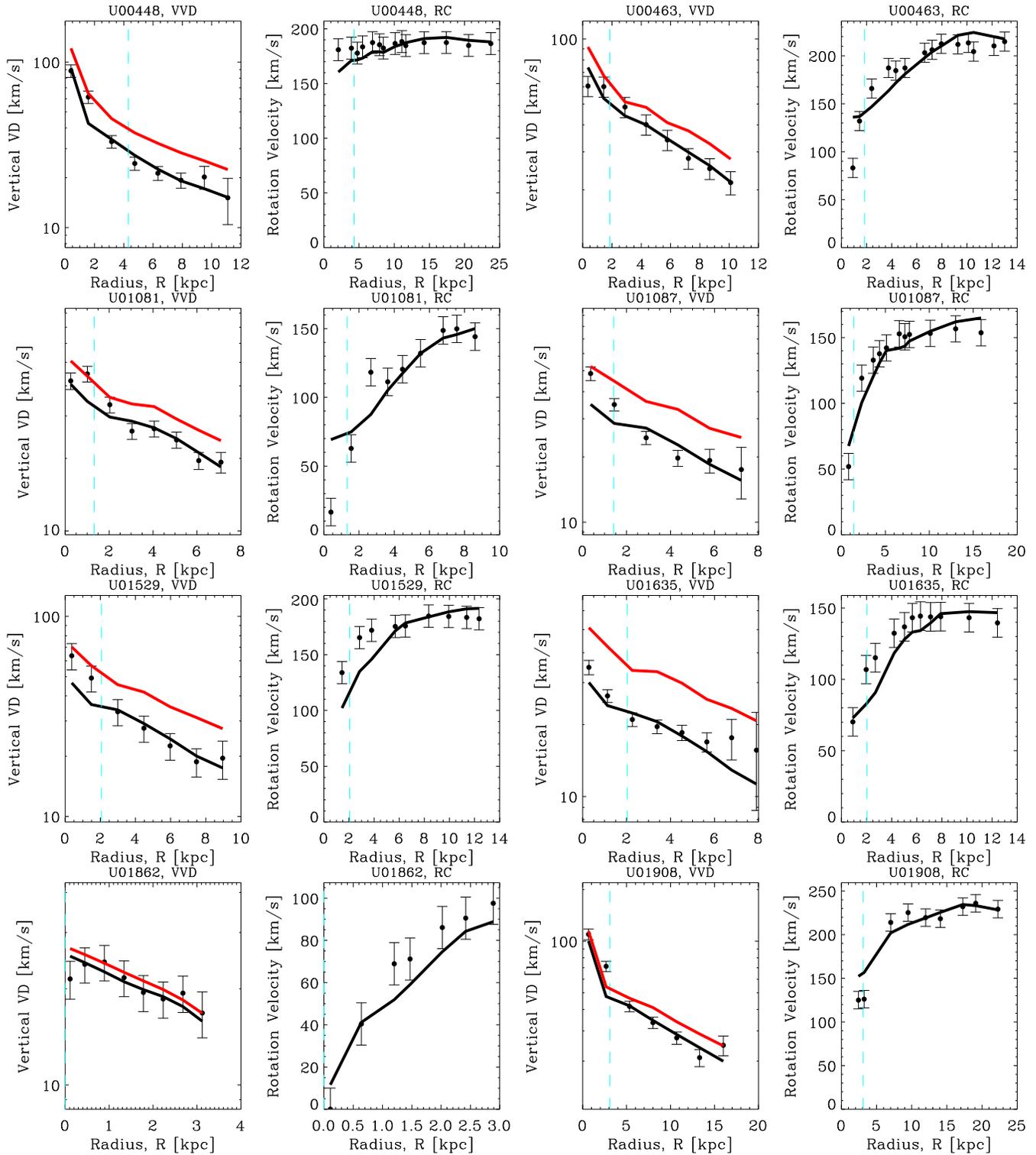}
\caption{Best fits (black lines) to the vertical velocity dispersions (first and third columns) and rotation curves (second and fourth columns) in MOND with $\gamma=1$. The vertical velocity dispersions found using the same best fit $M/L$ and inclination, but the scale-height derived from observations of edge-on galaxies (Eq~\ref{eqn:kreg}) are given by the red line. The data points are given in black and are derived from the raw measurements of DMSvi, but are corrected for the best fit inclination. The error bars on the vertical velocity dispersion data points are the combination of the systematic and random errors and the error bars on the rotation curve data points are 10~$\kms$ for each point as discussed in \S\ref{sec:errors}. The dashed turquoise vertical line shows the bulge radius, below which the vertical velocity dispersions are unreliable.}
\label{fig:3like0}
\end{figure*}

\begin{figure*}
\includegraphics[angle=0,width=18.0cm]{mondexfits.ps_pages1}
\caption{As per Fig~\ref{fig:3like0}.}
\label{fig:3like1}
\end{figure*}

\begin{figure*}
\includegraphics[angle=0,width=18.0cm]{mondexfits.ps_pages2}
\caption{As per Fig~\ref{fig:3like0}.}
\label{fig:3like2}
\end{figure*}

\begin{figure*}
\includegraphics[angle=0,width=18.0cm]{mondexfits.ps_pages3}
\caption{As per Fig~\ref{fig:3like0}.}
\label{fig:3like3}
\end{figure*}

\end{document}